\documentclass[12pt]{article}
\pdfoutput=1

\usepackage{cite}
\usepackage{amsmath}
\usepackage{amssymb}

\def\para{\\ [-2mm]}
\def \be  {\begin{equation}}
\def \ee  {\end{equation}}
\def \ba  {\begin{eqnarray}}
\def \ea  {\end{eqnarray}}
\newcommand{\nn}{\nonumber}
\def\eqn#1{eq.~(\ref{#1})} 
\def\Eqn#1{Equation~(\ref{#1})}
\def\eqns#1#2{eqs.~(\ref{#1}) and~(\ref{#2})}
\def\Eqns#1#2{Eqs.~(\ref{#1}) and~(\ref{#2})}

\def\IZ{\relax\ifmmode\mathchoice
{\hbox{\cmss Z\kern-.4em Z}}{\hbox{\cmss Z\kern-.4em Z}}
{\lower.4pt\hbox{\cmsss Z\kern-.4em Z}}
{\lower1.2pt\hbox{\cmsss Z\kern-.4em Z}}\else{\cmss Z\kern-.4em Z}\fi}
\newcommand{\Z}{\mathsf{Z}\kern -5pt \mathsf{Z}}
\newcommand{\unit}{\mathsf{1}\kern -3pt \mathsf{l}}
\def\ie{{i.e.}}
\def\viz{{viz.}}
\def\eg{{e.g.}}

\def\half{{\textstyle{1 \over 2}}}
\def\fr#1#2{ {\textstyle{#1 \over #2}}}

\def\de {\epsilon}

\def\eps{\varepsilon}
\def\cA {  {\cal A} }

\def\cL {  {\cal L}  }

\def\cO {  {\cal O} }
\def\< { \langle}
\def\> { \rangle}

\def\ta {\textsf{a}}
\def\tb {\textsf{b}}
\def\tc {\textsf{c}}
\def\td {\textsf{d}}
\def\te {\textsf{e}}

\def\tg{\textsf{g}}

\def\lam{\lambda}

\def\Ups {  \Upsilon }
\def\ups {  \upsilon }
\def\Th {  \Theta }
\def\th {  \theta} 

\def\tPsi { {\tilde \Psi} }
\def\tpsi { {\tilde \psi} }
\def\tPhi { {\tilde \Phi} }
\def\tphi { {\tilde \phi} }
\def\tTh { {\tilde \Th} }
\def\tth { {\tilde \th} }
\def\tUps { {\tilde \Ups} }
\def\tups {  {\tilde \ups} }
\def\tXi { {\tilde \Xi} }

\def\Tamn {\left(T^\ta\right)^{mn}}
\def\Ta {\left(T^\ta\right)}

\def\Tc {\left(T^\tc\right)}
\def\Td {\left(T^\td\right)}

\def\ms { m_s }
\def\mf { m_f }
\def\SYM { S_{\rm YM}^\mu }
\def\Smat { S_{\rm mat}^\mu }

\def\delslash{ \rlap{/}\partial }
\def\Aslash{ \rlap{\,/}A }
\def\Dslash{ \rlap{\,/}D }

\def\kslash{ \rlap{/}k }
\def\epsslash{ \rlap{/}{\eps}  }

\begin{document}

\titlepage
\begin{flushright}
BOW-PH-173 \\
\end{flushright}

\vspace{3mm}

\begin{center}
{\Large\bf\sf
Color-factor symmetry using perturbiner methods \\[2mm]
for tree-level amplitudes 
of Yang-Mills theory coupled to matter
}

\vskip 1.5cm

{\sc
Abigail R. Chriss$^{a,b}$\footnote{achriss@uchicago.edu}
\\[2mm]
Mia Karlsson$^a$\footnote{mkarlsson@bowdoin.edu} 
\\[2mm]
Stephen G. Naculich$^a$\footnote{naculich@bowdoin.edu}
}
\\

\vspace*{0.5cm} 
$^a$ {\it Department of Physics and Astronomy\\
Bowdoin College \\ 
Brunswick, ME 04011 USA\\
}

\vspace*{0.5cm} 
$^b$ {\it Department of Physics \\
University of Chicago\\
Chicago, IL 60637 USA\\
}
\vspace{5mm}

\begin{abstract}
Bern-Carrasco-Johansson relations are helicity-independent constraints
among the partial amplitudes of tree-level processes containing gluons and
possibly also matter fields.  They are a consequence of color-kinematic
duality, but alternatively can be derived using the color-factor symmetry
of tree-level amplitudes.  We use recursive perturbiner methods to
present a general, streamlined proof of the color-factor symmetry of
all tree-level amplitudes containing an arbitrary number of fundamental
spin-zero or spin-one-half matter fields and at least one gluon.
\end{abstract}

\vspace*{0.5cm}
\vfil\break

\end{center}

\section{Introduction}
\setcounter{equation}{0}
\label{sec:intro}

The Bern-Carrasco-Johansson (BCJ) relations are a set of linear
helicity-independent constraints among the partial amplitudes of
tree-level gauge theory processes involving at least one gluon.
These relations were discovered to be a consequence of color-kinematic
duality of Yang-Mills theory \cite{Bern:2008qj,Bern:2010ue,Bern:2010yg}
and subsequently of a much broader class of field theories (see
ref.~\cite{Bern:2019prr} for a review).  The subsequent proof of the
BCJ relations for all-gluon amplitudes using string-theory techniques
\cite{BjerrumBohr:2009rd,Stieberger:2009hq} and BCFW on-shell recursion
\cite{Feng:2010my,Chen:2011jxa} provided evidence for the conjecture of
tree-level color-kinematic duality.  Although not proven, the conjecture
that color-kinematic duality also applies to integrands of loop-level
amplitudes has been tested for amplitudes of various multiplicities
and loop levels in supersymmetric Yang-Mills theories.  Color-kinematic
duality has been particularly powerful in the construction of supergravity
amplitudes through the double-copy procedure \cite{Bern:2019prr}.  
\para

BCJ relations are also present for tree-level gauge theory
amplitudes involving an arbitrary number of scalar or spinor
fields transforming in the fundamental representation of the
gauge group, provided the amplitude contains at least one gluon
\cite{Zhu:1980sz,Goebel:1980es,Johansson:2014zca,Naculich:2014naa,Johansson:2015oia}.
Once again, these relations are a consequence of the assumption of
color-kinematic duality, and were subsequently proven using BCFW on-shell
recursion \cite{delaCruz:2015dpa}.
\para

R.~W.~Brown and one of current authors showed that BCJ relations could
alternatively be understood to follow from a property of tree-level gauge
theory amplitudes called {\it color-factor symmetry}, which acts as a
momentum-dependent shift on the color factors of an amplitude, leaving the
full amplitude invariant \cite{Brown:2016mrh,Brown:2016hck,Brown:2018wss}.
Color-factor symmetry was proven for both pure Yang-Mills theory and
for gauge theories with massive particles of various spins using the
{radiation vertex expansion} \cite{Brown:1982xx}.
\para

Color-kinematic duality and color-factor symmetry are closely related but
distinct features of gauge theories.  While the former implies the latter
(proven using the {cubic vertex expansion} in ref.\cite{Brown:2016mrh}),
the latter implies a less stringent (but gauge-invariant) constraint
than the former on the kinematic numerators of tree-level amplitudes
\cite{Brown:2016mrh}, yet is sufficient to yield the BCJ relations.
At loop level, color-kinematic duality implies color-factor symmetry,
but no independent proof of the latter has yet been developed.
\para

A new proof of tree-level color-factor symmetry for pure Yang-Mills
theory (as well as biadjoint scalar theory) was presented in
ref.~\cite{Naculich:2023wyp} using a perturbiner approach.
This alternative proof is more streamlined than the radiation
vertex expansion used in ref.~\cite{Brown:2016mrh}.  Moreover, the
recursive methods employed are likely more familiar to readers as
they are closely related to the Berends-Giele method for computing
tree-level QCD amplitudes.  Berends and Giele introduced a set of
partially off-shell amplitudes (known as Berends-Giele currents)
that could be computed recursively \cite{Berends:1987me}.  Rosly and
Selivanov \cite{Rosly:1996vr,Selivanov:1998hn,Selivanov:1999as}
subsequently showed that a perturbative solution of the classical
equations of motion (called the {\it perturbiner expansion}) acts as a
generating function for Berends-Giele currents.  Mafra, Schlotterer,
et al. \cite{Lee:2015upy,Mafra:2015vca,Mafra:2016ltu,Garozzo:2018uzj}
also used classical equations of motion to generate Berends-Giele
currents in various theories.  Mizera and Skrzypek \cite{Mizera:2018jbh}
introduced the {\it color-dressed perturbiner expansion}, which makes
the color factors in each term explicit: we find this form useful in the
demonstration of color-factor symmetry in ref.~\cite{Naculich:2023wyp}
and in this paper.
\para

The perturbiner expansion has had several important applications,
including the work of Lopez Arcos, Quintero V\'elez, et. al., who
related the $L_\infty$-algebra that appears in Batalin-Vilkovisky
quantization \cite{Macrelli:2019afx} to the perturbiner expansion for
biadjoint scalar and Yang-Mills theories \cite{Lopez-Arcos:2019hvg}
as well as in gauge theories with matter \cite{Gomez:2020vat}.
Berends-Giele currents in BCJ gauge were constructed using Bern-Kosower
rules \cite{Ahmadiniaz:2021fey}, with this work extended to gravity
using the double-copy procedure \cite{Ahmadiniaz:2021ayd}.  Gomez and
Jusinskas have applied perturbiner methods to gravity coupled to
matter \cite{Gomez:2021shh}, and in ref.~\cite{Armstrong:2022mfr},
perturbiner methods were used to compute tree-level boundary correlators
in anti-de Sitter space.  The perturbiner approach has also been found
effective for computing one-loop integrands \cite{Gomez:2022dzk}.
The connection between tree-level Berends-Giele recursion relations
and the $L_\infty$-algebra uncovered in ref.~\cite{Macrelli:2019afx}
was extended to loop-level recursion relations and the quantum
homotopy algebra $A_\infty$ in ref.~\cite{Jurco:2019yfd}.  For
connections between the homotopy algebra and the double copy, see
refs.~\cite{Borsten:2021hua,Escudero:2022zdz,Borsten:2022ouu}.
More recent applications of the perturbiner approach include
refs.~\cite{Cho:2021nim,Lee:2022aiu,Cho:2022faq,Cho:2023kux,Damgaard:2024fqj,
Tao:2025fch,Tao:2025pnt,Tao:2026irg}.
A comprehensive review of perturbiner methods has just appeared
in ref.~\cite{LipinskiJusinskas:2026ctz}.
\para

The purpose of this paper is to extend the perturbiner approach to
Yang-Mills theory coupled to scalar or spinor fields in the fundamental
representation (see also ref.~\cite{Gomez:2020vat}) and to use this
approach to prove color-factor symmetry for tree-level amplitudes
containing at least one gluon.  After developing recursion relations which
give the perturbative solutions to the nonlinear equations of motion
of the theory, and therefore yield all the tree-level amplitudes,
we prove that these amplitudes are invariant under color-factor
shifts, thus confirming in a more transparent way the results of
ref.~\cite{Brown:2016mrh,Brown:2016hck}.
\para

The outline of this paper is as follows.  In sec.~\ref{sec:bcj}, we
review the form of the BCJ relations for tree-level amplitudes containing
both gluons and fundamental matter fields.  In sec.~\ref{sec:cfs}, we
review how color-factor symmetry acts on amplitudes and show that BCJ
relations follow as a consequence.  In sec.~\ref{sec:pert}, we use the
classical equations following from the Yang-Mills plus matter Lagrangian
to develop the color-dressed perturbiner expansion for the calculation of
tree-level amplitudes.  In sec.~\ref{sec:proof},  we use the perturbiner
expansion to prove color-factor symmetry for all tree-level amplitudes
with an arbitrary number of fundamental matter fields and at least
one gluon.  In sec.~\ref{sec:concl}, we summarize our findings. In
appendix \ref{sec:app} we illustrate the use of the perturbiner approach
to compute four-point amplitudes for matter-antimatter annihilation for
both scalar and spinor fields, explicitly exhibiting color-kinematic
duality for these amplitudes.

\section{BCJ relations for matter-gluon amplitudes}
\setcounter{equation}{0}
\label{sec:bcj}

In this section, we review the form of the BCJ relations for 
amplitudes containing both gluons and matter fields 
(either scalar or spinor) 
transforming in the fundamental representation of the gauge group.
\para

We first illustrate the BCJ relations using one of the simplest
matter-gluon amplitudes, the four-point function for matter-antimatter
annihilation\footnote{In secs.~\ref{sec:bcj} and \ref{sec:cfs}, we use
$\Phi$ to denote either a scalar or a spinor matter field.}
$\bar{\Phi} \Phi \to g g$:
\begin{align}
\cA_4 (\bar{\Phi}_1, \Phi_2, g_3, g_4)
& = 
g^2 \left[  
 {n_{(1)} c_{(1)} \over 2 k_1 \cdot k_4} 
+{n_{(2)} c_{(2)} \over 2 k_2 \cdot k_4}
+{n_{(3)} c_{(3)} \over 2 k_3 \cdot k_4}
\right]  \,.
\label{fourpointamp}
\end{align}

Here $k_i$ are the four-momenta of the external particles, obeying
$k_1^2=k_2^2=m^2$ and $k_3^2 = k_4^2=0$  and momentum conservation
$\sum_{i=1}^4 k_i = 0$.
The color factors $c_{(i)}$ are defined as
\begin{align}
c_{(1)} = - (T^{\ta_4} T^{\ta_3})^{m_1 m_2} \,,\qquad
c_{(2)} =   (T^{\ta_3} T^{\ta_4})^{m_1 m_2} \,,\qquad
c_{(3)} =   f^{\ta_4 \ta_3 \te} (T^\te)^{m_1 m_2} \,.
\label{colorfactors}
\end{align}

In consequence of the fundamental representation commutation relations
\begin{align}
[T^\ta, T^\tb] = f^{\ta\tb\tc} T^\tc 
\label{commutationrelations}
\end{align}
the color factors (\ref{colorfactors}) obey the Jacobi identity
\begin{align}
c_{(1)} + c_{(2)} + c_{(3)} = 0 \,.
\label{colorjacobi}
\end{align}

For scalar matter fields,
the kinematic numerators $n_{(i)}$ are given by 
\begin{align}
n_{(1)} 
& = \eps_4 \cdot k_1 \,( -4 \eps_3 \cdot k_2 )
+\eps_4 \cdot \eps_3 \,(-2 k_1 \cdot k_4) \,,
\nn\\[2mm]
n_{(2)} 
& = \eps_4 \cdot k_2 \,( 4 \eps_3 \cdot k_1 )
+\eps_4 \cdot \eps_3 \,(2 k_2 \cdot k_4) \,,
\nn\\[2mm]
n_{(3)} 
& = \eps_4 \cdot k_1 \,(  4 \eps_3 \cdot k_2)
+\eps_4 \cdot k_2 \,( -4 \eps_3 \cdot k_1)
+\eps_4 \cdot \eps_3 \,(-2 k_2 \cdot k_4 + 2 k_1 \cdot k_4) 
\label{scalarnumerators}
\end{align}

whereas for spinor matter fields,
the kinematic numerators $n_{(i)}$ are given by 
\begin{align}
n_{(1)} 
& =   2 \eps_4 \cdot k_3 (\overline{v}_1  \epsslash_3 u_2)
    -   \overline{v}_1 \kslash_{3} \epsslash_4 \epsslash_3 u_2  
    +   2 \eps_3 \cdot k_2 (\overline{v}_1 \epsslash_4  u_2)  \,,
\nn\\[2mm]
n_{(2)} 
& =    \overline{v}_1 \kslash_{3} \epsslash_4   \epsslash_3 u_2
    -  2 \eps_4 \cdot  \eps_3 (\overline{v}_1 \kslash_{3}  u_2)
    +  2 \eps_3 \cdot k_1 (\overline{v}_1  \epsslash_4 u_2) \,,
\nn\\[2mm]
n_{(3)} 
& = 
   2 \eps_4 \cdot \eps_3 ( \overline{v}_{1} \kslash_3 u_2 )
- 2 \eps_4  \cdot k_3 (  \overline{v}_1 \epsslash_3  u_2 )
+ 2 \eps_3 \cdot k_4 (  \overline{v}_1 \epsslash_4  u_2 )
\label{spinornumerators}
\end{align}

where $\eps_i$ denotes the gluon polarization and $u$ and $\bar{v}$
denote initial state spinors.  \Eqns{scalarnumerators}{spinornumerators}
may be obtained either by evaluating the relevant Feynman diagrams, or by
using perturbiner methods, as illustrated in appendix \ref{sec:app}.  
\para

Upon inspection, one sees that the kinematic numerators 
for both scalar matter fields 
(\ref{scalarnumerators})
and spinor matter fields
(\ref{spinornumerators})
obey the kinematic Jacobi identity\footnote{In the case of the spinor fields,
$n_{(1)} + n_{(2)} + n_{(3)} 
= 2 \eps_3 \cdot (k_1 + k_2 + k_4) \overline{v}_1  \epsslash_4 u_2$.
This vanishes using $\sum_{i=1}^4 k_i = 0$ and $\eps_3 \cdot k_3=0$.}
\begin{align}
n_{(1)} + n_{(2)} + n_{(3)} = 0
\label{kinematicjacobi}
\end{align}
as was first recognized in refs.~\cite{Zhu:1980sz,Goebel:1980es}.
The analogy between \eqns{colorjacobi}{kinematicjacobi} has 
been subsequently dubbed color-kinematic duality.
\para

We  may use \eqn{colorjacobi} to express  \eqn{fourpointamp}
in terms of an independent basis of color factors
\begin{align}
\cA_4 (\bar{\Phi}_1, \Phi_2, g_3, g_4)
&= 
A(1,3,4,2) ~(T^{\ta_3} T^{\ta_4})^{m_1 m_2} 
+A(1,4,3,2)~  (T^{\ta_4} T^{\ta_3})^{m_1 m_2} 
\label{fourpointproper}
\end{align}

where the partial amplitudes $A(1,3,4,2)$ and $A(1,4,3,2)$ 
are given by
\begin{align} 
A(1,3,4,2) &=  g^2 \left( {n_{(2)} \over 2 k_2 \cdot k_4}
-{n_{(3)}  \over 2 k_3 \cdot k_4} \right) \,,
\nn\\
A(1,4,3,2) &= 
g^2 \left( - {n_{(1)} \over 2 k_1 \cdot k_4} 
+ {n_{(3)} \over 2 k_3 \cdot k_4} \right) \,.
\end{align}

One sees immediately that the partial amplitudes 
$A(1,3,4,2)$ and $A(1,4,3,2)$ obey a kinematic constraint 
\begin{align}
k_1 \cdot k_4  ~ A(1,4,3,2) - k_2 \cdot k_4  ~ A(1,3,4,2) = 0 
\label{bcjfour}
\end{align}
as a consequence of \eqn{kinematicjacobi}, 
together with  the analogous constraint on denominators
\begin{align}
k_1 \cdot k_4 + k_2 \cdot k_4 + k_3 \cdot k_4=0
\end{align}
which follows from momentum conservation and $k_4^2=0$. 
Equation (\ref{bcjfour}) is the simplest example of a ``BCJ relation.''
\para

Bern, Carrasco, and Johansson \cite{Bern:2008qj}
discovered that the kinematic numerators of 
five- and higher-point all-gluon tree-level amplitudes 
could be chosen (using a generalized gauge transformation)
to obey Jacobi relations mirroring those satisfied by the color factors
(i.e., color-kinematic duality).
As a result, they derived relations 
analogous to \eqn{bcjfour} for the partial amplitudes 
(defined using a trace basis) of $n$-gluon amplitudes.
These BCJ relations were subsequently proven through a variety of approaches
\cite{BjerrumBohr:2009rd,Stieberger:2009hq,Feng:2010my,Chen:2011jxa},
thus validating the assumption of color-kinematic duality at tree level.
\para

BCJ relations are also obeyed by amplitudes containing 
a pair of fundamental matter particles and an arbitrary number of gluons, 
as was shown in ref.~\cite{Naculich:2014naa}
using the assumption of color-kinematic duality
together with momentum conservation.
Specifically, when such amplitudes are expressed 
using a proper decomposition\cite{Kosower:1987ic,Mangano:1988kk},
\ie, one in which the color factors form an independent basis,
such as 
\begin{align}
\cA_{n} (\bar{\Phi}_1, \Phi_2, g_3, \cdots, g_n)
~=~ \sum_{\gamma \in S_{n-2}}  
A (1, \gamma(3), \cdots, \gamma(n),2)
~\left( {T}^{\ta_{\gamma(3)}}{T}^{\ta_{\gamma(4)}}
\cdots {T}^{\ta_{\gamma(n)}} \right)^{m_1 m_2}
\label{proper}
\end{align}

with $\gamma$ a permutation of $\{3, \cdots, n\}$,
then the partial amplitudes $A(1, \gamma(3), \cdots, \gamma(n), 2)$
satisfy the constraint
\begin{align}
\sum_{b=3}^{n} \left( k_n \cdot k_1 
+ \sum_{c=3}^{b-1} k_n \cdot k_{\sigma(c)}  \right) 
A (1, \sigma(3), \cdots, \sigma(b-1), n, \sigma(b), \cdots, \sigma(n-1),2)  ~=~0
\label{bcj}
\end{align}

where $\sigma$ is an arbitrary 
permutation of $\{3, \cdots, n-1\} $.
Known as the fundamental BCJ 
relation \cite{BjerrumBohr:2009rd,Feng:2010my,Sondergaard:2011iv},
this is the $n$-point generalization of \eqn{bcjfour}.
These relations were explicitly verified in ref.~\cite{Naculich:2014naa}
for various five- and six-point amplitudes
using known results \cite{Bern:1996ja,Badger:2005zh,Forde:2005ue},
providing evidence for the assumption of color-kinematic duality 
for this class of amplitudes. 
\para

Finding a proper decomposition is a more subtle problem
when there are more than two fundamental matter fields in the amplitude, 
but was solved by Melia \cite{Melia:2013bta,Melia:2013epa}
and Johansson and Ochirov \cite{Johansson:2015oia}.
Johansson and Ochirov then demonstrated 
(again assuming color-kinematic duality)
that the primitive amplitudes $A(1, \gamma(3), \cdots, \gamma(n),2)$
in the Melia-Johansson-Ochirov decomposition of amplitudes 
containing at least one gluon obey relations precisely of the form (\ref{bcj}),
where $n$ represents one of the gluons.
(See refs.~\cite{Johansson:2015oia,Brown:2016hck}
for a precise characterization of these primitive amplitudes.)
These BCJ relations were subsequently proven 
by de la Cruz, Kniss, and Weinzierl \cite{delaCruz:2015dpa}
for all $n$ using BCFW on-shell recursion,
again validating the assumption of color-kinematic duality
at tree level.
\para

In sec.~\ref{sec:cfs} we will review how the BCJ relations (\ref{bcj})
can alternatively be derived using the invariance of the amplitudes
under a color-factor shift.

\section{Color-factor symmetry of matter-gluon amplitudes}
\setcounter{equation}{0}
\label{sec:cfs}

In 2016, a new angle on BCJ relations was found when R.~W.~Brown and one
of the current authors discovered a new symmetry (\viz, color-factor
symmetry) of tree-level amplitudes containing at least one gluon.
This symmetry was proven for amplitudes with an arbitrary number
of gluons as well as fundamental matter fields of various spins in
refs.~\cite{Brown:2016mrh,Brown:2016hck} using  the radiation vertex
expansion.  In ref.~\cite{Naculich:2023wyp}, an alternative proof
using perturbiner methods was presented for all-gluon amplitudes.
In secs.~\ref{sec:pert} and \ref{sec:proof} of this paper, this
perturbiner proof will be extended to amplitudes with an arbitrary number
of scalar or spinor fundamentals  and at least one gluon.
\para

In this section, we illustrate how
invariance of a tree-level amplitude 
under color-factor symmetry leads directly to the BCJ relations
without assuming color-kinematic duality.
We first give a brief reprise of color-factor symmetry, 
referring to 
refs.~\cite{Brown:2016mrh,Brown:2016hck,Brown:2018wss,Naculich:2023wyp}
for more details.
\para

Tree-level scattering amplitudes of a gauge theory 
can be expressed as a decomposition \cite{Bern:2010tq}
\begin{align}
\cA_n =\sum_i a_{(i)}  c_{(i)} 
\label{colordecomposition}
\end{align}

where $c_{(i)} $ are color factors consisting of the contraction of
various color tensors $f^{\ta\tb\tc}$ and $(T^\ta)^{mn}$ appearing in
the Feynman diagrams, and $a_{(i)} $ depends on kinematic and spin
factors, \eg,  see \eqn{fourpointamp}.  There exists a color-factor
symmetry associated with each external gluon $a$ contributing to the
amplitude \cite{Brown:2016mrh,Brown:2016hck}.  This symmetry acts on
each color factor $c_{(i)} $ appearing in \eqn{colordecomposition}
by a momentum-dependent shift $\delta_a c_{(i)} $.  The simplest way to
characterize the shift is in terms of color tensors appearing in $c_{(i)}$
\cite{Cheung:2021zvb,Naculich:2023wyp}
\begin{align}
\delta_a f^{\tb \ta \tc} &=  \alpha_a \delta^{\tb\tc} (k_c^2 - k_b^2) \,,
\nn\\[2mm]
\delta_a(T^\ta)^{mn} &= \alpha_a\delta^{mn}(k^2_n-k^2_m) 
\label{shift}
\end{align}
where $\alpha_a$ is a constant and $k_i$ denote the four-momenta flowing
out of the other two legs associated with the vertex.  
\para

Now consider the proper decomposition of the amplitude with one pair of
fundamental matter particles and $n-2$ gluons (\ref{proper}).  A generic
term in the sum can be written as
\begin{align}
A (1, \sigma(3), \cdots, \sigma(b-1), n, \sigma(b), \cdots, \sigma(n-1),2) 
~
\left( {T}^{\ta_{\sigma(3)}} \cdots {T}^{\ta_{\sigma(b-1)}}
T^{\ta_n} {T}^{\ta_{\sigma(b)}}\cdots {T}^{\ta_{\sigma(n-1)}} \right)^{m_1 m_2}
\end{align}
for $\sigma$ an arbitrary permutation of $\{3, \cdots, n-1\} $
with $n$ shuffled in between $\sigma(b-1)$ and $\sigma(b)$.
The color-factor shift associated with the $n$th gluon
causes the product of generators to transform as
\begin{align}
\delta_{n} 
\left( {T}^{\ta_{\sigma(3)}} \cdots {T}^{\ta_{\sigma(b-1)}}
T^{\ta_n} {T}^{\ta_{\sigma(b)}}\cdots {T}^{\ta_{\sigma(n-1)}} \right)^{m_1 m_2}
= 
\alpha_n 
\left( {T}^{\ta_{\sigma(3)}} \cdots {T}^{\ta_{\sigma(n-1)}} \right)^{m_1 m_2}
\left(  K_2^2 - K_1^2 \right)
\label{traceshift}
\end{align}
where
$K_1 = k_1 + \sum_{c=3}^{b-1} k_{\sigma(c)} $
and
$K_2 = k_2  + \sum_{c=b}^{n-1} k_{\sigma(c)} $.
Using $k_n^2=0$ together with momentum conservation $K_1 + K_2 + k_n=0$, one has
\begin{align}
K_2^2 - K_1^2 = 
2 \left( k_n \cdot k_1 
+ \sum_{c=3}^{b-1} k_n \cdot k_{\sigma(c)}  \right)  \,.
\label{K1K2}
\end{align}

We now invoke the fact,
already proven in ref.~\cite{Brown:2016mrh}
and verified using perturbiner methods in the remainder 
of this paper, 
that \eqn{proper} is invariant under the color 
factor shift associated with the $n$th gluon
\begin{align}
\delta_n 
\cA_{n} (\bar{\Phi}_1, \Phi_2, g_3, \cdots, g_n) = 0  \,.
\end{align}

Applying \eqns{traceshift}{K1K2} to \eqn{proper}, 
and recognizing that the products of generators 
$\left({T}^{\ta_{\sigma(3)}}\cdots {T}^{\ta_{\sigma(n-1)}}\right)^{m_1 m_2}$ 
are linearly independent, one directly establishes the fundamental BCJ
relations (\ref{bcj}) for an amplitude with one pair of fundamental
matter particles and $n-2$ gluons.
\para

As already discussed in sec.~\ref{sec:bcj}, 
tree-level amplitudes containing an arbitrary number
of fundamental matter fields and an arbitrary number of gluons
can be expressed in terms of primitive amplitudes 
in the Melia-Johansson-Ochirov proper decomposition.
In ref.~\cite{Brown:2016hck}, it was shown 
by applying \eqn{shift} to the MJO decomposition 
and invoking color-factor symmetry 
that the primitive amplitudes 
obey the $n$-point BCJ relations (\ref{bcj}),
provided the amplitude contains at least one gluon,
as was previously found in ref.~\cite{Johansson:2015oia}.
\para

Thus in all cases, one may replace the logic 
\begin{align}
\hbox{color-kinematic duality} \longrightarrow \hbox{BCJ relations}
\nn
\end{align}
with 
\begin{align}
\hbox{color-factor symmetry} \longrightarrow \hbox{BCJ relations} \,.
\nn
\end{align}

In the case of the four-point amplitude, color-factor symmetry leads
directly to color-kinematic duality, as follows.  \Eqn{shift} implies
that the color factors $c_{(i)}$ appearing in the four-point amplitude
(\ref{fourpointamp}) undergo shifts
\begin{align}
\delta_{4} c_{(1)} 
&= \alpha_4 (T^{\ta_3})^{m_1 m_2} (-2 k_1 \cdot k_4)  \,,
\nn\\
\delta_{4} c_{(2)}
&= \alpha_4 (T^{\ta_3})^{m_1 m_2}(- 2 k_1 \cdot k_3 ) \,,
\nn\\
\delta_{4} c_{(3)} 
&= \alpha_4 (T^{\ta_3})^{m_1 m_2} (-2 k_3 \cdot k_4 )
\end{align}

so that 
\begin{align}
\delta_4 \cA_4 (\bar{\Phi}_1, \Phi_2, g_3, g_4)
& = 
- g^2 \alpha_4 (T^{\ta_3})^{m_1 m_2} 
\left[  n_{(1)} +n_{(2)} +n_{(3)} \right] \,.
\end{align}

The invariance of the four-point amplitude 
$\delta_4 \cA_4 (\bar{\Phi}_1, \Phi_2, g_3, g_4)=0$
immediately implies the kinematic Jacobi identity (\ref{kinematicjacobi}).
\para

In general, however, color-factor symmetry does not directly
imply color-kinematic duality, but rather leads to constraints
\cite{Brown:2016mrh,Brown:2016hck} on the kinematic numerators that are
less stringent than the kinematic Jacobi relations.\footnote{For example
see refs.~\cite{BjerrumBohr:2010zs,Tye:2010dd} for constraints for the
five-gluon amplitude.}
These constraints, however, are gauge-invariant 
(whereas the kinematic Jacobi relations only hold in
a particular generalized gauge) and are sufficient to yield the BCJ
relations, as we have seen.
\para

\section{Perturbiner approach to tree-level amplitudes}
\setcounter{equation}{0}
\label{sec:pert}

In this section, we review and extend the perturbiner approach
to compute tree-level gauge theory amplitudes 
containing an arbitrary number of scalar or spinor fields 
transforming in the fundamental representation
together with an arbitrary number of gluons.
The results of this section will be used in sec.~\ref{sec:proof}
to provide a proof of the invariance of these amplitudes under
a color factor shift, 
a more streamlined alternative to the original
proof given in ref.~\cite{Brown:2016mrh}.
\para

First, we recast the equations of motion following from the
Yang-Mills-matter Lagrangian into a form useful for the perturbiner
computation.  Second, we derive the solutions of the linearized equations.
Third, we develop the recursive equations of the perturbiner approach,
and finally we use these to obtain an expression for the $n$-point
tree-level amplitude.

\subsection{Equations of motion}

The Lagrangian for the Yang-Mills field coupled to spinor and 
scalar matter is 
\begin{align}
\cL &= 
-\fr{1}{4} F^\ta_{\mu\nu} F^{\ta\mu\nu}  
 \, + \, \overline \Psi (i \Dslash - \mf  ) \Psi
\, + \, (D^\mu \Phi)^\dag  (D_\mu \Phi) - \ms^2 \Phi^\dag \Phi
\label{lagrangian}
\end{align}

where 
\begin{align}
F^{\nu\mu \ta} &= 
\partial^\nu A^{\mu \ta} -\partial^\mu A^{\nu \ta} 
-i g f^{\ta\tb\tc} A^{\nu \tb} A^{\mu \tc} \,, 
\nn \\[2mm]
D_\mu \Psi &= (\partial_\mu - i g A^\ta_\mu T^\ta )\Psi \,,
\nn \\[2mm]
D_\mu \Phi &= (\partial_\mu - i g A^\ta_\mu T^\ta )\Phi 
\label{defFD}
\end{align}

and the generators in the defining representation 
satisfy \eqn{commutationrelations}.
The Euler-Lagrange equations following from \eqn{lagrangian} are 
\begin{align}
D_\nu F^{\nu\mu \ta}
&= - g  \overline\Psi \gamma^\mu T^\ta \Psi 
+ i g  
\left[ (D^\mu \Phi)^\dag T^\ta \Phi - \Phi^\dag T^\ta D^\mu \Phi \right]  
\,,
\nn \\[2mm]
(-i \Dslash + \mf  ) \Psi &= 0 \,,
\nn \\[2mm]
(D_\mu D^\mu + \ms^2)  \Phi 
&= 0 \,.
\label{eom}
\end{align}

Using \eqn{defFD}, one has
\begin{align}
D_\nu F^{\nu\mu \ta}
&= \partial_\nu F^{\nu\mu \ta} -i g f^{\ta\tb\tc} A^\tb_\nu F^{\nu\mu \tc}  
\nn\\
&= 
\partial^2 A^{\mu \ta} - \partial^\mu (\partial_\nu A^{\nu \ta} )
- i g f^{\ta\tb\tc} (\partial_\nu A^{\nu \tb} ) A^{\mu \tc}
- i g f^{\ta\tb\tc} A^\tb_\nu (\partial^\nu A^{\mu \tc} + F^{\nu\mu \tc})   \,,
\nn\\[3mm]
D_\mu D^\mu \Phi 
&= \partial^2 \Phi 
- ig (\partial_\mu A^{\mu\ta}) T^\ta \Phi 
- ig A_{\mu}^\ta T^\ta (\partial^\mu \Phi +  D^\mu \Phi ) \,.
\end{align}

Imposing Lorenz gauge 
\begin{align}
\partial_\mu A^{\mu \ta} = 0
\label{lorenz}
\end{align}

one can rewrite the equations of motion (\ref{eom}) as 
\begin{align}
-\partial^2 A^{\mu \ta} &= 
  g f^{\ta\tb\tc} A^\tb_\nu G^{\nu\mu \tc}  
+ g  \overline\Psi \gamma^\mu T^\ta \Psi  
- g  \left( \Th^{\mu\dag} T^\ta \Phi + \Phi^\dag T^\ta \Th^\mu \right)  \,,
\nn\\[2mm]
(-i \delslash + \mf ) \Psi &=  g \Aslash^{\ta} T^\ta \Psi \,,
\nn\\[2mm]
\overline{\Psi} (i \overleftarrow{ \rlap{/}\partial } + \mf )
&=  g    \overline{\Psi} \Aslash^{\ta} T^\ta \,,
\nn\\[2mm]
-(\partial^2 + \ms^2) \Phi &=   g A_{\mu}^\ta T^\ta \Upsilon^\mu
\label{eom2}
\end{align}

where $\overline{\Psi} = \Psi^\dag \gamma^0$ 
is the Dirac conjugate spinor field,
and we have defined several auxiliary fields
\begin{align}
G^{\nu\mu \ta} 
&\equiv
-i (\partial^\nu A^{\mu \ta} + F^{\nu\mu \ta} ) 
= -i (2 \partial^\nu A^{\mu \ta} 
-\partial^\mu A^{\nu \ta}) 
 -  g f^{\ta\tb\tc} A^{\nu \tb} A^{\mu \tc}  \,,
\nn\\[2mm]
\Th^\mu 
& \equiv  
-i D^\mu \Phi 
= -i \partial^\mu \Phi -  g A^{\mu\ta} T^\ta \Phi \,, 
\nn \\[2mm]
\Upsilon^\mu 
&\equiv 
-i (\partial^\mu \Phi +  D^\mu \Phi ) 
 = -2 i \partial^\mu \Phi - g A^{\mu\ta} T^\ta\Phi \,.
\label{defaux}
\end{align}

The advantage of introducing these auxiliary fields is that
the equations of motion (\ref{eom2}) remain at most quadratic 
(rather than cubic) in all the fields,
greatly simplifying their application to the perturbiner ansatz, 
as recognized by Mafra, Schlotterer, et. al
\cite{Lee:2015upy,Mafra:2015vca,Mafra:2016ltu,Garozzo:2018uzj,Mizera:2018jbh}.
\para

Finally we deduce equations of motion for the auxiliary fields (\ref{defaux}) 
using \eqn{eom2}:
\begin{align}
-\partial^2 G^{\nu\mu \ta} 
&= 
  i \left[ 2 \partial^\nu \partial^2 A^{\mu \ta} 
-\partial^\mu \partial^2 A^{\nu \ta} - i g f^{\ta\tb\tc}\partial^2  (A^{\nu \tb} A^{\mu \tc} ) \right]
\nn\\
&= 
- i g f^{\ta\tb\tc} \left[ 2 \partial^\nu ( A^\tb_\lambda G^{\lambda\mu \tc} )
                 - \partial^\mu  ( A^\tb_\lambda G^{\lambda\nu \tc} ) 
                 +i \partial^2 ( A^{\nu \tb} A^{\mu \tc} ) \right]
\nn  \\
& 
~~~~
-2ig \partial^\nu (  \overline\Psi \gamma^\mu T^\ta \Psi )
+ig  \partial^\mu (  \overline\Psi \gamma^\nu T^\ta \Psi  )
\nn  \\
& 
~~~~~+ i g \left[
2 \partial^\nu \left(\Th^{\mu\dag} T^\ta \Phi + \Phi^\dag T^\ta \Th^\mu \right) 
-\partial^\mu  \left( \Th^{\nu\dag} T^\ta \Phi + \Phi^\dag T^\ta \Th^\nu \right) 
\right] \,,
\nn\\[3mm]
-(\partial^2 + \ms^2) \Th^\mu 
&=  i \partial^\mu (\partial^2 + \ms^2) \Phi
   +g (\partial^2 + \ms^2) (A^{\mu\ta} T^\ta \Phi) \nn\\
&= -i g \partial^\mu  ( A_{\nu}^\ta T^\ta \Upsilon^\nu )
   +g (\partial^2 + \ms^2) (A^{\mu\ta} T^\ta \Phi) \,,
\nn
\\[3mm]
-(\partial^2 + \ms^2) \Upsilon^\mu 
&=  2 i \partial^\mu (\partial^2 + \ms^2) \Phi
   +g (\partial^2 + \ms^2) (A^{\mu\ta} T^\ta \Phi) \nn\\
&= -2 i g \partial^\mu  ( A_{\nu}^\ta T^\ta \Upsilon^\nu )
   +g (\partial^2 + \ms^2) (A^{\mu\ta} T^\ta \Phi) \,.
\label{eomaux}
\end{align}

\subsection{Linearized solutions}

Next we solve the linearized equations of motion 
(\ie, dropping all $\cO(g)$ terms).
Equations (\ref{eom2}) reduce to 
\begin{align}
\partial^2 A^{\mu \ta} = 0 \,,
\qquad
(-i \delslash + \mf ) \Psi^m =0 \,,
\qquad
\overline{\Psi} (i \overleftarrow{ \rlap{/}\partial } + \mf ) \,,
\qquad
(\partial^2 + \ms^2) \Phi^n =   0
\end{align}

and have solutions that are superpositions of plane waves
\begin{align}
A^{\mu \ta} (x) 
&=  \sum_i 
(A_i^{\mu \ta}  e^{-i k_i \cdot x} 
+A_i^{\mu \ta*}  e^{i k_i \cdot x}) + \cO(g) 
&\hbox{where}\qquad 
k_i^2 &=0  \,,
\nn\\
\Psi^n(x) 
&= \sum_i 
(\Psi_i^n   e^{-i k_i \cdot x} 
+ \tPsi_i^n   e^{i k_i \cdot x}) + \cO(g)
&\hbox{where}\qquad 
k_i^2 &=\mf^2  \,,
\nn\\
\overline{\Psi}^{n} (x) 
&= \sum_i
(\overline{\tPsi}_i^{n} e^{ -i k_i \cdot x}
+\overline{\Psi}_i^{n}  e^{ i k_i \cdot x}) + \cO(g)
&\hbox{where}\qquad 
k_i^2 &=\mf^2  \,,
\nn\\
\Phi^n(x) 
&= \sum_i 
(\Phi_i^n   e^{-i k_i \cdot x} 
+ \tPhi_i^n   e^{i k_i \cdot x}) + \cO(g) 
&\hbox{where}\qquad 
k_i^2 &=\ms^2 
\end{align}

with (using the Lorenz condition (\ref{lorenz}))
\begin{align}
A_i^{\mu \,\ta}  &= \eps_i^\mu \delta^{\ta\ta_i}
&\hbox{where}~~\qquad\qquad \eps_i \cdot k_i &= 0  \,,
\nn\\[2mm]
\Psi_i^n &= \psi_i \delta^{nn_i}
&\hbox{where}\qquad
(\kslash_i - \mf) \psi_i &= 0 \,, 
\nn\\[2mm]
\overline{\tPsi}_i^n &= \overline{\tpsi}_i \delta^{nn_i}
&\hbox{where}\qquad
\overline{\tpsi}_i ( \kslash_i + \mf )  &= 0 \,,
\nn\\[2mm]
\Phi_i^n &= \phi_i \delta^{nn_i}  \,,
\nn\\[2mm]
\tPhi_i^{n} &= \tphi_i \delta^{nn_i} \,.
\label{linear}
\end{align}

Equations (\ref{eomaux}) for the auxiliary fields reduce to 
\begin{align}
\partial^2 G^{\nu\mu \ta} 
= 0\,, \qquad
(\partial^2 + \ms^2) \Th^{\mu  n}
= 0\,, \qquad
(\partial^2 + \ms^2) \Upsilon^{\mu  n}
= 0
\end{align}

and are solved by 
\begin{align} 
G^{\nu\mu \ta} (x) 
&= \sum_i 
(G_i^{\nu\mu \ta} e^{-i k_i \cdot x} 
+G_i^{\nu\mu \ta*} e^{i k_i \cdot x} ) + \cO(g) 
&\hbox{where}\qquad 
k_i^2 &=0 \,,
\nn\\
\Th^{\mu n} (x) 
&= \sum_i 
(\Th_i^{\mu n}   e^{-i k_i \cdot x} 
+ \tTh_i^{\mu n}   e^{i k_i \cdot x}) + \cO(g)  
&\hbox{where}\qquad 
k_i^2 &=\ms^2  \,,
\nn\\
\Ups^{\mu n} (x) 
&= \sum_i 
(\Ups_i^{\mu n}   e^{-i k_i \cdot x} 
+ \tUps_i^{\mu n}   e^{i k_i \cdot x}) + \cO(g)  
&\hbox{where}\qquad 
k_i^2 &=\ms^2  \,.
\end{align}

Using the linearized versions of eqs.~(\ref{defaux})
\begin{align}
G^{\nu\mu \ta} 
 = -i (2 \partial^\nu A^{\mu \ta} -\partial^\mu A^{\nu \ta} )+  \cO(g)  \,,
\qquad
\Th^\mu 
 =
 -i \partial^\mu \Phi +\cO(g)  \,,
\qquad
\Upsilon^\mu 
 = -2 i \partial^\mu \Phi +\cO(g)
\end{align}

one obtains
\begin{align}
G_i^{\nu\mu\,\ta} &= g_i^{\nu\mu} \delta^{\ta\ta_i}
&\hbox{where} \qquad
g_i^{\nu\mu} &= -2 k_i^\nu \eps_i^\mu + k_i^\mu \eps_i^\nu  \,,
\nn\\[2mm]
\Th_i^{\mu n} &= \th_i^\mu \delta^{nn_i} 
&\hbox{where} \qquad 
\th_i^\mu &=  -k_i^\mu \phi_i  \,,
\nn\\[2mm]
\tTh_i^{\mu n} &= \tth_i^{\mu} \delta^{nn_i} 
&\hbox{where}\qquad 
\tth_i^{\mu} &= k_i^\mu \tphi_i \,,
\nn\\[2mm]
\Ups_i^{\mu n} &= \ups_i^\mu \delta^{nn_i} 
&\hbox{where} \qquad 
\ups_i^\mu &= -2k_i^\mu \phi_i  \,,
\nn\\[2mm]
\tUps_i^{\mu n} &= \tups_i^{\mu} \delta^{nn_i} 
&\hbox{where} \qquad 
\tups_i^{\mu} &= 2k_i^\mu \tphi_i \,.
\label{linearaux}
\end{align}

\subsection{Perturbiner expansion}

The linearized solutions obtained in the previous section are the first
terms in a perturbative solution of the nonlinear equations of motion,
known as the perturbiner ansatz. 
This ansatz, introduced by Rosly and Selivanov
\cite{Rosly:1996vr,Selivanov:1998hn,Selivanov:1999as},
consists of an infinite series of plane waves:
\begin{align}
A^{\mu \ta} (x) 
&= \sum_P 
(A_P^{\mu \ta}  e^{-i k_P \cdot x}
+A_P^{\mu \ta*}  e^{i k_P \cdot x})
\,,\nn\\
\Psi^n (x) 
&= \sum_P
(\Psi^n_P   e^{-i k_P \cdot x} 
+ \tPsi^n_P   e^{i k_P \cdot x})
\,,\nn\\
\overline{\Psi}^{n} (x) 
&= \sum_P
(\overline{\tPsi}_P^{n}   e^{ -i k_P \cdot x}
+\overline{\Psi}_P^{n}   e^{i k_P \cdot x})
\,,\nn\\
\Phi^n (x) 
&= \sum_P 
(\Phi^n_P   e^{-i k_P \cdot x} 
+ \tPhi^n_P   e^{i k_P \cdot x})
\,,\nn\\
G^{\nu\mu \ta} (x) 
&= \sum_P 
(G_P^{\nu\mu \ta} e^{-i k_P \cdot x} 
+G_P^{\nu\mu \ta*} e^{i k_P \cdot x} )
\,,\nn\\
\Th^{\mu n} (x) 
&= \sum_P 
(\Th_P^{\mu n}   e^{-i k_P \cdot x} 
+ \tTh_P^{\mu n}   e^{i k_P \cdot x})
\,,\nn\\
\Ups^{\mu n} (x) 
&= \sum_P 
(\Ups_P^{\mu n}   e^{-i k_P \cdot x} 
+ \tUps_P^{\mu n}   e^{i k_P \cdot x})
\label{perturbineransatz}
\end{align}

where the sum runs over all non-empty ordered words 
$P=p_1 p_2 \cdots p_m$ with $p_1 < p_2 < \cdots < p_m$
and $k_P = \sum_{j=1}^m k_{p_j}$.
We are using a version of the perturbiner ansatz developed by Mizera
and Skrzypek \cite{Mizera:2018jbh}, called the {\it color-dressed
perturbiner expansion} (to be distinguished from the color-stripped
perturbiner expansion) in which the color indices are shown explicitly.
The linearized solutions correspond to one-letter words, which are used
to generate the coefficients with multi-letter words using the nonlinear
equations of motion.  Plugging \eqn{perturbineransatz} into the equations
of motion (\ref{eom2}) and (\ref{eomaux}), one obtains

\begin{align}
k_P^2 A_P^{\mu  \ta} 
&=  g f^{\ta\tb\tc} \sum_{P=Q \cup R} 
A_{\nu Q}^{\tb} G_R^{\nu\mu \tc}  
+ g (T^\ta)^{mn} \sum_{P=Q \cup R}
\overline{\tPsi}_{Q}^m \gamma^\mu  \Psi^n_R
\nn\\
& \hspace{44mm}
- g (T^\ta)^{mn} \sum_{P=Q \cup R} 
\left[ {\tTh}^{\mu m * }_{Q} \Phi^n_R
+ {\tPhi}^{m *}_{Q}  \Th^{\mu n}_R \right] \,,
\label{kA} \\[2mm]
( \kslash_P - \mf ) \Psi_P^m
&= - g (T^\ta)^{mn} \sum_{P=Q \cup R}  \Aslash_Q^{ \ta} \Psi_R^n \,,
\label{kPsi} \\[2mm]
\overline{\tPsi}_P^m
( \kslash_P + \mf )
&=  g (T^\ta)^{nm} \sum_{P=Q \cup R} \overline{\tPsi}_Q^n \Aslash_R^{ \ta} \,,
\label{kPsitilde} \\[2mm]
(k_P^2 - \ms^2) \Phi_P^m 
&= g (T^\ta)^{mn} \sum_{P=Q \cup R}  A_{\mu Q}^{\ta} \Ups^{\mu n}_R \,,
\label{kPhi}\\[2mm]
(k_P^2 - \ms^2) \tPhi_P^{m  *}
&= g (T^\ta)^{nm} \sum_{P=Q \cup R}  
\tUps^{\mu n *}_Q  A_{\mu R}^{\ta}  
\label{kPhitilde}
\end{align}

for the gauge and matter fields, and 
\begin{align}
k_P^2 G_P^{\nu\mu  \ta}
&= g f^{\ta\tb\tc} \sum_{P=Q \cup R}
\left[- 2k_P^\nu A_{\lam Q}^{\tb} G_R^{\lam\mu \tc}
+k_P^\mu A_{\lam Q}^{\tb} G_R^{\lam\nu \tc}
-k_P^2 A_Q^{\nu  \tb} A_R^{\mu \tc} \right] 
\nn
\\
& ~~~~ + g (T^\ta)^{mn} \sum_{P=Q \cup R}
\left[
-2 k^\nu_P ~\overline{\tPsi}^m_Q \gamma^\mu \Psi_R^n
+ k^\mu_P ~\overline{\tPsi}^m_Q \gamma^\nu \Psi_R^n
\right]
\label{kG}
\\
& ~~~~ + g (T^\ta)^{mn} \sum_{P=Q \cup R} 
\left[ 2 k^\nu_P (  {\tTh}^{\mu m*}_Q \Phi_R^n + {\tPhi}^{m*}_Q \Th_R^{\mu n} )
- k^\mu_P (  {\tTh}^{\nu m*}_Q \Phi_R^n + {\tPhi}^{m*}_Q \Th_R^{\nu n} ) \right] \,,
\nn\\[2mm]
(k_P^2 - \ms^2) \Th_P^{\mu m}
&= g (T^\ta)^{mn} \sum_{P=Q \cup R} 
\left[- k^\mu_P A^{\ta}_{\nu Q} \Ups^{\nu n}_R
 - (k_P^2-\ms^2)  A^{\mu \ta}_{Q} \Phi^n_R \right] \,,
\label{kTh}\\[2mm]
(k_P^2 - \ms^2) \tTh_P^{\mu m *}
&= g (T^\ta)^{nm} \sum_{P=Q \cup R} 
\left[ k^\mu_P \tUps^{\nu n *}_Q A^{\ta}_{\nu R} 
- (k_P^2-\ms^2) \tPhi^{n*}_Q  A^{\mu \ta}_{R} \right] \,,
\label{kThtilde}\\[2mm]
(k_P^2 - \ms^2) \Ups_P^{\mu m}
&= g (T^\ta)^{mn} \sum_{P=Q \cup R} 
\left[ -2 k^\mu_P A^{\ta}_{\nu Q} \Ups^{\nu n}_R
 - (k_P^2-\ms^2)  A^{\mu \ta}_{Q} \Phi^n_R \right] \,,
\label{kUps}\\[2mm]
(k_P^2 - \ms^2) \tUps_P^{\mu m *}
&= g (T^\ta)^{nm} \sum_{P=Q \cup R} 
\left[ 2 k^\mu_P \tUps^{\nu n *}_Q A^{\ta}_{\nu R} 
- (k_P^2-\ms^2) \tPhi^{n*}_Q  A^{\mu \ta}_{R} \right]
\label{kUpstilde}
\end{align}

for the auxiliary fields, 
where $P = Q \cup R$ denotes all possible divisions of $P$ into 
two non-empty ordered words $Q$ and $R$.
Here we emphasize that using the auxiliary fields (\ref{defaux})
in addition to $A^{\mu \ta} $, $\Psi^n$,  and $\Phi^n$ 
makes eqs.~(\ref{kA})-(\ref{kUpstilde}) at most quadratic in fields. 
Applying the Lorenz gauge condition (\ref{lorenz})
to \eqn{perturbineransatz} implies
\begin{align}
k_P^\mu A_{\mu P}^\ta &=0
\label{lorenzP}
\end{align}

and inserting \eqn{perturbineransatz} 
into the definitions of the auxiliary fields (\ref{defaux})
yields the following constraints 
\begin{align}
G^{\nu\mu\ta}_P
&= -2k^\nu_PA^{\mu\ta}_P+k^\mu_PA^{\nu\ta}_P-H^{\nu\mu\ta}_P \,,
\label{G}
\\[2mm]
\Th^{\mu n}_P&=-k^\mu_P\Phi^n_P-\Xi^{\mu n}_P  \,,
\label{Th}
\\[2mm]
\tTh^{\mu n*}_P&=k^\mu_P\tPhi^{n*}_P-\tXi^{\mu n*}_P  \,,
\label{Thtilde}
\\[2mm]
\Ups^{\mu n}_P&=-2k^\mu_P\Phi^n_P-\Xi^{\mu n}_P \,,
\label{Ups}
\\[2mm]
\tUps^{\mu n*}_P&=2k^\mu_P\tPhi^{n*}_P-\tXi^{\mu n*}_P
\label{Upstilde}
\end{align}

where for later convenience we have introduced 
\begin{align}
H^{\nu\mu\ta}_P
&=gf^{\ta\tb\tc}\sum_{P=Q\cup R}A^{\nu\tb}_QA^{\mu\tc}_R  \,,
\label{defH}
\\
\Xi^{\mu m}_P
&=g\Tamn \sum_{P=Q\cup R}A^{\mu\ta}_Q\Phi^n_R  \,,
\label{defXi}
\\
\tXi^{\mu m*}_P
&=g\Ta^{nm}\sum_{P=Q\cup R}\tPhi^{n*}_Q  A^{\mu\ta}_R  \,.
\label{defXitilde}
\end{align}

\subsection{Evaluation of amplitudes}

The coefficients $A^{\mu \,\ta}_P$ in the perturbiner ansatz 
(\ref{perturbineransatz}) 
are the (color-dressed) Berends-Giele currents of 
the Yang-Mills theory \cite{Berends:1987me}.
To obtain a tree-level $n$-point amplitude containing a gluon,
one first computes $A_P^{\mu\ta}$ for $P=12\cdots (n-1)$,
where $1, \cdots, n-1$ denote all the other fields in the amplitude.
Since momentum conservation for the $n$-point amplitude 
implies $k_P = - k_n$, and an on-shell amplitude has $k_n^2=0$,
one extracts the residue of the $k_P^2$ pole.
Finally one contracts
with the Berends-Giele current $A_{\mu n}^{\ta}$
of the remaining gluon to obtain \cite{Berends:1987me,Mizera:2018jbh}
\begin{align}
\cA_n=\lim_{k_P^2 \to 0} A^{\ta}_{\mu{n}} k^2_P A^{\mu\ta}_P
\qquad\hbox{with}\qquad P=123{\dots}(n-1)  \,.
\label{amp}
\end{align}

Using \eqns{linear}{kA}, one may rewrite the amplitude as 
\begin{align}
\cA_n
=
g \eps_{\mu n} \Bigg( f^{\ta_n\tb\tc}
\sum_{P=Q\cup{R}}A^\tb_{\nu{Q}}G^{\nu\mu\tc}_R
&+  (T^{\ta_n})^{mn} \sum_{P=Q \cup R}
\overline{\tPsi}_{Q}^m \gamma^\mu  \Psi^n_R
\nn\\
&
-  (T^{\ta_n})^{mn}\sum_{P=Q\cup{R}}
\left[\tTh^{\mu{m*}}_Q\Phi^n_R +\tPhi^{m*}_Q\Th^{\mu{n}}_R\right]
\Bigg)   \,.
\label{npointamp}
\end{align}
We illustrate the use of \eqn{npointamp} in appendix \ref{sec:app} by
explicitly evaluating the amplitude for matter-antimatter
annihilation into two gluons.
\Eqn{npointamp}
is also the starting point for our proof of color-factor symmetry
in sec.~\ref{sec:proof}.

\section{Proof of color-factor symmetry for Yang-Mills $+$ matter}
\setcounter{equation}{0}
\label{sec:proof}

The goal of this section is to use the recursive perturbiner expansion
to confirm that any tree-level amplitude containing an arbitrary number
of scalar or spinor matter fields and at least one gluon is invariant
under the color-factor shift associated with that gluon.
We begin with the expression (\ref{npointamp}) for the $n$-point
amplitude, where $n$ denotes a gluon and all the other fields may be
either gluons, scalars, or spinor fields.
We apply the color-factor shift (\ref{shift}) to obtain
\begin{align}
\delta_n\cA_n&=g\alpha_n\de_{\mu n}\sum_{P=Q\cup{R}}\delta^{\tb\tc}(k^2_Q-k^2_R)A^\tb_{\nu Q}G^{\nu\mu\tc}_R 
\nn \\
&~~~~
-g\alpha_n\de_{\mu n}\sum_{P=Q\cup{R}}\delta^{mn}(k^2_Q-k^2_R)
\overline{\tPsi}_{Q}^m \gamma^\mu  \Psi^n_R
\nn \\
&~~~~
+g\alpha_n\de_{\mu n}\sum_{P=Q\cup{R}}\delta^{mn}(k^2_Q-k^2_R)
\left(
\tPhi^{m*}_Q\Th^{\mu{n}}_R + \tTh^{\mu{m}*}_Q\Phi^n_R
\right) 
\end{align}

where $P=123{\dots}(n-1) $.
We now rewrite this as 
\begin{align}
\delta_n\cA_n&=g\alpha_n\de_{\mu n}S^\mu 
\label{dAmp}
\end{align}

with
\begin{align}
S^\mu 
=\sum_{P=Q\cup{R}}\Big[
&
(k^2_QA^{\tc}_{\nu{Q}})G^{\nu\mu\tc}_R
-A^{\ta}_{\nu{Q}}(k^2_RG^{\nu\mu\ta}_R)
\nn\\
&
-\overline{\tPsi}_{Q}^n (\kslash_Q+\mf)(\kslash_Q-\mf) \gamma^\mu  \Psi^n_R
+\overline{\tPsi}_{Q}^m \gamma^\mu (\kslash_R+\mf)(\kslash_R-\mf) \Psi^m_R
\nn\\[2mm]
&
+(k^2_Q-\ms^2)\tPhi^{n*}_Q\Th^{\mu{n}}_R
-\tTh^{\mu{m}*}_Q(k^2_R-\ms^2)\Phi^m_R
\nn\\[2mm]
&
+(k^2_Q-\ms^2)\tTh^{\mu{n}*}_Q\Phi^n_R 
-\tPhi^{m*}_Q(k^2_R-\ms^2)\Th^{\mu{m}}_R 
\Big]
\label{sMu}.
\end{align}

The goal is now to prove that $S^\mu$ vanishes when 
contracted with $\eps_{\mu n}$.
We use eqs.~(\ref{kA})-(\ref{kThtilde}) 
and relabel the indices to obtain:
\begin{align}
S^\mu&=
gf^{\ta\tb\tc}\sum_{P=A\cup B\cup C}
\left[A^\ta_{\nu A}G^{\nu\tb}_{~\lam B}G^{\lam\mu\tc}_C
+A^\ta_{\nu A}\left(
 2k^\nu_{BC}A^\tb_{\lam B}G^{\lam\mu\tc}_C
-k^\mu_{BC}A^\tb_{\lam B}G^{\lam\mu\tc}_C
+k^2_{BC}A^{\nu\tb}_BA^{\mu\tc}_C \nn \right) \right]
\\
&+g\Tamn \sum_{P=A\cup B\cup C}
\Big[
\overline{\tPsi}_{B}^m \gamma_\nu  \Psi^n_C G^{\nu\mu\ta}_A
+A^{\ta}_{\nu{A}}
\left(
 2 k^\nu_{BC}
  \overline{\tPsi}^m_B \gamma^\mu \Psi_C^n
- k^\mu_{BC}
  \overline{\tPsi}^m_B \gamma^\nu \Psi_C^n
\right)
\nn \\
&\hspace{34mm}
- \left(\tTh_{\nu B}^{m*}\Phi^n_C +\tPhi^{m*}_B\Th_{\nu C}^{n}\right)
   G^{\nu\mu\ta}_{A}
\nn\\[2mm]
&\hspace{34mm} 
-  2A^\ta_{\nu A}k^\nu_{BC}
   \left(\tTh^{\mu m*}_B\Phi^n_C +\tPhi^{m*}_B\Th^{\mu n}_C\right) 
+A^\ta_{\nu A}k^\mu_{BC}
\left(\tTh^{\nu m*}_B\Phi^n_C +\tPhi^{m*}_B\Th^{\nu n}_C\right)
\nn\\[2mm]
&\hspace{34mm}
-\overline{\tPsi}_{B}^m \Aslash_A^\ta (\kslash_{AB}-\mf) \gamma^\mu  \Psi^n_C
-\overline{\tPsi}_{B}^m \gamma^\mu (\kslash_{AC} +\mf) \Aslash_A^\ta  \Psi^n_C
\nn\\[2mm]
&\hspace{34mm}
+\tUps^{\nu m*}_BA^\ta_{\nu A}\Th^{\mu n}_C 
-\tTh^{\mu m*}_BA^\ta_{\nu A}\Ups^{\nu n}_C
\nn \\[2mm]
&\hspace{34mm} 
+k^\mu_{AB}\tUps^{\nu m*}_BA^\ta_{\nu A}\Phi^n_C 
+k^\mu_{AC}\tPhi^{m*}_BA^\ta_{\nu A}\Ups^{\nu n}_C
\nn \\[2mm]
&\hspace{34mm} 
-(k^2_{AB}-\ms^2)\tPhi^{m*}_BA^{\mu\ta}_A\Phi^n_C
+(k^2_{AC}-\ms^2)\tPhi^{m*}_BA^{\mu\ta}_A\Phi^n_C\Big] 
\label{twosums}
\end{align}

where $k_{AB}^\mu$ denotes $ k_A^\mu + k_B^\mu$, etc.
We express \eqn{twosums} as a sum of contributions
\begin{align}
S^\mu = \SYM + \Smat 
\end{align}

where $\SYM$ is the contribution that arises in a pure Yang-Mills
theory and $\Smat$ contains all the additional terms that arise from
the matter terms.
The pure Yang-Mills term 
was computed in ref.~\cite{Naculich:2023wyp} to be\footnote{The sign change 
in the second term is due to the change in convention from 
$A_P^{\mu \ta}  e^{ i k_P \cdot x}$ to 
$A_P^{\mu \ta}  e^{-i k_P \cdot x}$ in the perturbiner ansatz.}
\begin{align}
\SYM 
 &=
k_P^\mu 
\bigg[ 
g f^{\ta\tb\tc} \sum_{P=A\cup B\cup C}
 k_A \cdot A_C^\tc \ A_A^\ta \cdot A_B^\tb
-
\half g^2  
f^{\ta\tb\te}
f^{\te\tc\td}
\sum_{P=A\cup B\cup C \cup D}
A_A^\ta \cdot A_C^\tc
\ A_B^\tb \cdot A_D^\td 
\bigg] 
\label{SYM}
\end{align}

and arises from the first sum in \eqn{twosums},
namely those terms containing $f^{\ta\tb\tc}$. 
In the Yang-Mill plus matter theory considered in this paper,
the first sum gives rise to an additional contribution.
Equation (4.26) of ref. \cite{Naculich:2023wyp} 
contains a term of the form
\begin{align}
gf^{\td\tc\ta}\sum_{A=D\cup I\cup A}
A^\td_{\nu D}\left(k^2_IA^{\nu\tc}_I\right)A^{\mu\ta}_A \,.
\end{align}

We see that the matter fields in \eqn{kA} of this paper 
give rise to new terms, namely
\begin{align}
S^\mu_1
&= g^2f^{\ta\td\tc}\Tc^{m n}\sum_{P=A\cup B\cup C\cup D}
\left[
\overline{\tPsi}_{B}^m \gamma^\nu  \Psi^n_C
- \tTh^{\nu m*}_B\Phi^n_C-\tPhi^{m*}_B\Th^{\nu n}_C
\right]
A^\td_{\nu D}A^{\mu\ta}_A 
\label{S1} \\
&= g^2 f^{\ta\td\tc}\Tc^{m n}\sum_{P=A\cup B\cup C\cup D}
\left[
\overline{\tPsi}_{B}^m \gamma^\nu  \Psi^n_C
+ (k^\nu_C-k^\nu_B) \tPhi^{m*}_B\Phi^n_C
+\tXi^{\nu m*}_B\Phi^n_C+\tPhi^{m*}_B\Xi^{\nu n}_C
\right]
A^\td_{\nu D}A^{\mu\ta}_A
\nn
\end{align}
where we have applied \eqns{Th}{Thtilde} in the second line.
\para

We now split the second sum in \eqn{twosums},
namely those terms containing $\Tamn$, 
into nine terms $S^\mu_2$ through $S^\mu_{10}$,
defined in eqs.~(\ref{S2})-(\ref{S10}) below, 
so that the matter field contribution is given in its entirety by
\begin{align}
\Smat = \sum_{i=1}^{10} S_i^\mu \,.
\end{align}

We now define and simplify $S^\mu_2$ through $S^\mu_{10}$, 
repeatedly using eqs.~(\ref{G})-(\ref{Upstilde}).
First
\begin{align}
S_2^\mu
&= g\Tamn \sum_{P=A\cup B\cup C}
\overline{\tPsi}_{B}^m \gamma_\nu  \Psi^n_C ~G^{\nu\mu\ta}_A
\nn\\
&= g\Tamn \sum_{P=A\cup B\cup C}
\overline{\tPsi}_{B}^m 
\left[ -2\kslash_A  A^{\mu\ta}_A + k^\mu_A \Aslash^{\ta}_A \right] 
\Psi^n_C
- \overline{\tPsi}_{B}^m \gamma_\nu  \Psi^n_C H^{\nu\mu\ta}_A \,.
\label{S2}
\end{align}

Next
\begin{align}
S_3^\mu
&= g\Tamn \sum_{P=A\cup B\cup C}
\overline{\tPsi}^m_B
\left(  2 k^\nu_{BC}  \gamma^\mu A^{\ta}_{\nu{A}}
- k^\mu_{BC}  \gamma^\nu  A^{\ta}_{\nu{A}} \right)
\Psi_C^n
\nn\\
&= g\Tamn \sum_{P=A\cup B\cup C}
\overline{\tPsi}^m_B
\left[ 2 \gamma^\mu (k_B+k_C) \cdot A^{\ta}_A
- (k_B^\mu + k_C^\mu) \Aslash^{\ta}_A \right]
\Psi_C^n \,.
\end{align}

We also have
\begin{align}
S_4^\mu
&= g\Tamn \sum_{P=A\cup B\cup C}\left(
-\tTh^{m*}_{\nu B}\Phi^n_C-\tPhi^{m*}_B\Th^{n}_{\nu C}\right)
G^{\nu\mu\ta}_{A} 
\\
&=g\Tamn \sum_{P=A\cup B\cup C}
\Big\{
\left[(-k_{\nu B}+k_{\nu C})\tPhi^{m*}_B\Phi^n_C\right]
\left(-2k^\nu_{A}A^{\mu\ta}_A+k^\mu_AA^{\nu\ta}_{A}\right) 
-\left[(-k_{\nu B}+k_{\nu C})\tPhi^{m*}_B\Phi^n_C\right]
H^{\nu\mu\ta}_{A}
\nn \\
&~~~~\hspace{30mm} 
+\left(\tXi^{m*}_{\nu B}\Phi^n_C+\tPhi^{m*}_B\Xi^{n}_{\nu C}\right)
\left(-2k^\nu_{A}A^{\mu\ta}_A+k^\mu_AA^{\nu\ta}_{A}\right) 
-\left(\tXi^{m*}_{\nu B}\Phi^n_C+\tPhi^{m*}_B\Xi^{n}_{\nu C}\right)
H^{\nu\mu\ta}_{A}
\Big\}.
\nn
\end{align}

Next we have
\begin{align}
S_5^\mu
&=g\Tamn \sum_{P=A\cup B\cup C}
\left[ -2k^\nu_{BC}
\left(\tTh^{\mu m*}_B\Phi^n_C+\tPhi^{m*}_B\Th^{\mu n}_C\right)\right]
A^\ta_{\nu A} \nn \\
&=g\Tamn \sum_{P=A\cup B\cup C}
2(k^\nu_B+k^\nu_C)
\left[(-k^\mu_B+k^\mu_C)\tPhi^{m*}_B\Phi^n_C
+\tXi^{\mu m*}_B\Phi^n_C+\tPhi^{m*}_B\Xi^{\mu n}_C\right]A^\ta_{\nu A}.
\end{align}

Also
\begin{align}
S_6^\mu
&= g\Tamn \sum_{P=A\cup B\cup C}k^\mu_{BC}
\left( \tTh^{\nu m*}_B\Phi^n_C+\tPhi^{m*}_B\Th^{\nu n}_C\right)
A^\ta_{\nu A} \nn
\\&=g\Tamn \sum_{P=A\cup B\cup C}(k^\mu_B+k^\mu_C)
\left[(k^\nu_B-k^\nu_C)\tPhi^{m*}_B\Phi^n_C
-\tXi^{\nu m*}_B\Phi^n_C-\tPhi^{m*}_B\Xi^{\nu n}_C\right]A^\ta_{\nu A}.
\end{align}

Continuing 
\begin{align}
S_7^\mu&=
g\Tamn \sum_{P=A\cup B\cup C}
\overline{\tPsi}_{B}^m \left[
-\Aslash_A^\ta ( \kslash_{AB}-\mf) \gamma^\mu
-\gamma^\mu (\kslash_{AC} +\mf) \Aslash_A^\ta
\right] \Psi^n_C
\nn\\
&=
 g\Tamn \sum_{P=A\cup B\cup C}
\overline{\tPsi}_{B}^m \Big[
- 2 \gamma^\mu k_A \cdot A_A^\ta
- 2 k_A^\mu \Aslash_A^\ta
+ 2 \kslash_A A_A^{\mu\ta}
- 2 \gamma^\mu (k_B + k_C) \cdot A_A^\ta
\nn\\
&~~~~~~\hspace{34mm}
+ (\kslash_B + \mf) \Aslash_A^\ta \gamma^\mu
+  \gamma^\mu \Aslash_A^\ta (\kslash_C - \mf)
\Big] \Psi^n_C
\nn\\[3mm]
&=
g\Tamn \sum_{P=A\cup B\cup C}
\overline{\tPsi}_{B}^m \Big[
- 2 k_A^\mu \Aslash_A^\ta
+ 2 \kslash_A A_A^{\mu\ta}
- 2 \gamma^\mu (k_B + k_C) \cdot A^\ta
\Big] \Psi^n_C
\nn\\
&~~~~+ g^2 \sum_{P=A\cup B\cup C \cup D}
\overline{\tPsi}_{B}^m \Big[
(T^\td T^\ta)^{mn}  \Aslash_D^\td \Aslash_A^\ta \gamma^\mu
-  (T^\ta T^\td)^{mn}  \gamma^\mu \Aslash_A^\ta \Aslash_D^\td
\Big] \Psi^n_C
\nn\\[3mm]
&=
g\Tamn \sum_{P=A\cup B\cup C}
\overline{\tPsi}_{B}^m \Big[
- 2 k_A^\mu \Aslash_A^\ta
+ 2 \kslash_A A_A^{\mu\ta}
- 2 \gamma^\mu (k_B + k_C) \cdot A^\ta
\Big] \Psi^n_C
\\
&~~~~+ g^2 \sum_{P=A\cup B\cup C \cup D}
\overline{\tPsi}_{B}^m \Big[
2 [T^\td, T^\ta]^{mn} \Aslash_D^\td A_A^{\mu\ta}
- (T^\td T^\ta)^{mn}  \Aslash_D^\td \gamma^\mu \Aslash_A^\ta
+  (T^\ta T^\td)^{mn}  \Aslash_A^\ta \gamma^\mu \Aslash_D^\td
\Big] \Psi^n_C
\nn
\end{align}

where we have made liberal use of
$\{\gamma^\mu, \gamma^\nu \} = 2 \eta^{\mu\nu} $ throughout,
and we have used eqs.~(\ref{kPsi}), (\ref{kPsitilde}), and (\ref{lorenzP})
in going from the second equation to the third. 
The last term in the last equation cancels
the penultimate term  upon relabelling
$ A \leftrightarrow B$.
\para

Next we have
\begin{align}
S_8^\mu
&=g\Tamn \sum_{P=A\cup B\cup C}
\left(-\tTh^{\mu m*}_B\Ups^{\nu n}_C
+\tUps^{\nu m*}_B\Th^{\mu n}_C\right)A^\ta_{\nu A} \nn
\\
&=g\Tamn \sum_{P=A\cup B\cup C}
\left[\left( k^\mu_B\tPhi^{m*}_B -\tXi^{\mu m*}_B\right)
\left(2k^\nu_C\Phi^n_C+\Xi^{\nu n}_C\right) 
-\left( 2k^\nu_B\tPhi^{m*}_B-\tXi^{\nu m*}_B\right)
\left(k^\mu_C\Phi^n_C+\Xi^{\mu n}_C\right)\right]A^\ta_{\nu A} \nn
\\
&=g\Tamn \sum_{P=A\cup B\cup C}
\Big[2(k^\nu_Ck^\mu_B-k^\nu_Bk^\mu_C)\tPhi^{m*}_B\Phi^n_C
+k^\mu_B\tPhi^{m*}_B\Xi^{\nu n}_C
+k^\mu_C\tXi^{\nu m*}_B\Phi^n_C
\nn \\
&~~~~ \hspace{34mm} 
-2k^\nu_C\tXi^{\mu m*}_B\Phi^n_C
-2k^\nu_B\tPhi^{m*}_B\Xi^{\mu n}_C
-\tXi^{\mu m*}_B\Xi^{\nu n}_C
+\tXi^{\nu m*}_B\Xi^{\mu n}_C\Big] A^\ta_{\nu A}.
\end{align}

Also
\begin{align}
S_9^\mu
&= g\Tamn \sum_{P=A\cup B\cup C}
\left( k^\mu_{AB}\tUps^{\nu m*}_B\Phi^n_C
+k^\mu_{AC}\tPhi^{m*}_B\Ups^{\nu n}_C\right)A^\ta_{\nu A} \nn\\
&=g\Tamn \sum_{P=A\cup B\cup C}
\left[2(k^\mu_{AB}k^\nu_B-k^\mu_{AC}k^\nu_C)\tPhi^{m*}_B\Phi^n_C
-k^\mu_{AB}\tXi^{\nu m*}_B\Phi^n_C
-k^\mu_{AC}\tPhi^{m*}_B\Xi^{\nu n}_C\right]A^\ta_{\nu A}
\end{align}

and finally, using \eqns{kPhi}{kPhitilde}, we obtain
\begin{align}
S_{10}^\mu
&=g\Tamn \sum_{P=A\cup B\cup C}
\left[-(k^2_{AB}-\ms^2)
+(k^2_{AC}-\ms^2)\right]\tPhi^{m*}_B\Phi^n_CA^{\mu\ta}_A \nn
\\
&=g\Tamn \sum_{P=A\cup B\cup C}
\left[-2k_A\cdot k_B+2k_A\cdot k_C
-(k^2_B-\ms^2)+(k^2_C-\ms^2)\right]\tPhi^{m*}_BA^{\mu\ta}_A\Phi^n_C \nn
\\
&=g\Tamn \sum_{P=A\cup B\cup C\cup D}
\big[
2k_A\cdot(k_C-k_B)\tPhi^{m*}_B\Phi^n_C
\nn
\\&~~~~ \hspace{38mm} 
-g\Td^{lm}\tUps^{\nu l*}_BA^\td_{\nu D}\Phi^n_C
+g\Td^{nl}\tPhi^{m*}_B A^\td_{\nu D}\Ups^{\nu l}_C\big]A^{\mu\ta}_A 
\nn \\
&=g\Tamn \sum_{P=A\cup B\cup C}
2k_A\cdot(k_C-k_B)\tPhi^{m*}_B\Phi^n_CA^{\mu\ta}_A \nn \\
&~~~~+g^2\sum_{P=A\cup B\cup C\cup D}
\Big\{
\left[-2(T^\td T^\ta)^{mn}k^\nu_B
-2(T^\ta T^\td)^{mn}k^\nu_C\right]\tPhi^{m*}_B\Phi^n_C 
\nn \\
&~~~~\hspace{32mm} 
+(T^\td T^\ta)^{mn}\tXi^{\nu m*}_B\Phi^n_C
-(T^\ta T^\td)^{mn}\tPhi^{m*}_B\Xi^{\nu n}_C 
\Big\}
A^\td_{\nu D}A^{\mu\ta}_A.
\label{S10}
\end{align}

We now collect all the results for $S_1^\mu$ through $S_{10}^\mu$
and redistribute them into three new terms 
\begin{align}
\Smat =
\sum_{i=1}^{10} S_i^\mu = S_{(1)}^\mu + S_{(2)}^\mu + S_{(3)}^\mu
\end{align}

where  $S_{(n)}^\mu$ consists of all terms containing $n$ powers of $g$.
For convenience, we further split $S_{(2)}$ into 
$S_{(2,1)}^\mu$, which consists of terms with the $\mu$ index on a momentum,
and $S_{(2,2)}^\mu$, which consists of terms with the $\mu$ index on a field. 
\para

We begin by collecting all the terms in eqs.~(\ref{S1})-(\ref{S10}) 
with one power of $g$:
\begin{align}
S^\mu_{(1)}
&=g\Tamn \sum_{P=A\cup B\cup C}\Big[
2(k_B-k_C)\cdot k_A \eta^{\mu\nu}	
+(k^\nu_C-k^\nu_B)k^\mu_A		
+2(k^\nu_B+k^\nu_C)(k^\mu_C-k^\mu_B)	
\nn \\
&~~~~ \hspace{34mm} 
+(k^\mu_B+k^\mu_C)(k^\nu_B-k^\nu_C) 	
+2(k^\nu_Ck^\mu_B-k^\nu_Bk^\mu_C)	
+2(k^\mu_A+k^\mu_B)k^\nu_B		
\nn\\[2mm]
&~~~~\hspace{34mm} 
-2(k^\mu_A+k^\mu_C)k^\nu_C		
+2k_A\cdot(k_C-k_B)\eta^{\mu\nu}	
\Big] \tPhi^{m*}_B\Phi^n_CA^\ta_{\nu A} 	
\nn \\[2mm]
&~~~~+g\Tamn\sum_{P=A\cup B\cup C}		
\overline{\tPsi}_{B}^m
\Big[
- 2\kslash_A  A^{\mu\ta}_A 
+ k^\mu_A \Aslash^{\ta}_A  			
+ 2 \gamma^\mu (k_B+k_C) \cdot A^{\ta}_A  
- (k_B^\mu + k_C^\mu) \Aslash^{\ta}_A    
\nn\\
&  \hspace{52mm} 
- 2 k_A^\mu \Aslash_A^\ta  
+ 2 \kslash_A A_A^{\mu\ta} 			
- 2 \gamma^\mu (k_B + k_C) \cdot A_A^\ta	
\Big] \Psi^n_C
\nn\\[2mm]
&=g\Tamn \sum_{P=A\cup B\cup C}
(k^\mu_A+k^\mu_B+k^\mu_C)
\left[ 
(k^\nu_B-k^\nu_C)
\tPhi^{m*}_B\Phi^n_CA^\ta_{\nu A} 
- \overline{\tPsi}_{B}^m \Aslash_A^\ta \Psi^n_C \right]
\nn
\\
&=g \Tamn
k^\mu_P 
\sum_{P=A\cup B\cup C}
\left[ (k_B-k_C) \cdot A^\ta_{A} \tPhi^{m*}_B\Phi^n_C
- \overline{\tPsi}_{B}^m \Aslash_A^\ta \Psi^n_C
\right] 
\label{firstorder}
\end{align}

where $k_P = k_A + k_B + k_C$ since $P=A\cup B\cup C$.
\para

Next, we collect all the terms in eqs.~(\ref{S1})-(\ref{S10}) 
with two powers of $g$ and in which the index $\mu$ labels a four-momentum:
\begin{align}
S^\mu_{(2,1)}
&=g\Tamn \sum_{P=A\cup B\cup C}
\Big[
 k^\mu_A\left(\tXi^{\nu m*}_B\Phi^n_C +\tPhi^{m*}_B\Xi^{\nu n}_C\right) 
-(k^\mu_B+k^\mu_C)
\left(\tXi^{\nu m*}_B\Phi^n_C+\tPhi^{m*}_B\Xi^{\nu n}_C\right) 
\nn \\
&~~~~ \hspace{34mm} 
+ k^\mu_B\tPhi^{m*}_B\Xi^{\nu n}_C 
+ k^\mu_C\tXi^{\nu m*}_B\Phi^n_C 
- (k^\mu_A+k^\mu_B)\tXi^{\nu m*}_B\Phi^n_C  
\nn \\[2mm]
&~~~~ \hspace{34mm} 
-(k^\mu_A+k^\mu_C)\tPhi^{m*}_B\Xi^{\nu n}_C\Big]A^\ta_{\nu A}  
\nn \\
&=g\Tamn \sum_{P=A\cup B\cup C}
\left(
-2k^\mu_B\tXi^{\nu m*}_B\Phi^n_C
-2k^\mu_C\tPhi^{m*}_B\Xi^{\nu n}_C\right)A^\ta_{\nu A} \nn
\\
&=-2g\Tamn \sum_{P=A\cup I\cup J}
\left(
 k^\mu_I\tXi^{\nu m*}_I\Phi^n_J
+k^\mu_J\tPhi^{m*}_I\Xi^{\nu n}_J\right)A^\ta_{\nu A}
\end{align}

where in the last line we have relabeled the indices 
to make the following steps clearer. 
We use \eqns{defXi}{defXitilde},
where $I = B \cup D$ in the first term (so that $k_I=k_B+k_D$)
and $J = C \cup D$ in the second term (so that $k_J=k_C+k_D$),
giving 
\begin{align}
S^\mu_{(2,1)}
&=-2g^2\sum_{P=A\cup B\cup C\cup D}
\Big[ (T^\td T^\ta)^{mn} (k^\mu_B+k^\mu_D)
+(T^\ta T^\td)^{mn} (k^\mu_C+k^\mu_D)\Big]
\tPhi^{m*}_B\Phi^n_C  A^\ta_{\nu A} A^{\nu\td}_D \,.
\end{align}

Relabeling $\ta\leftrightarrow\td$ and $A\leftrightarrow D$  
in the first term, we obtain
\begin{align}
S^\mu_{(2,1)}
&=-2g^2\sum_{P=A\cup B\cup C\cup D}
(T^\ta T^\td)^{mn}(k^\mu_B+k^\mu_A+k^\mu_C+k^\mu_D)
\tPhi^{m*}_B\Phi^n_C A^\ta_{\nu A} A^{\nu\td}_D
\nn \\
&=k_P^\mu \bigg[ -2g^2 \sum_{P=A\cup B\cup C\cup D}
(T^\ta T^\td)^{mn}\tPhi^{m*}_B\Phi^n_C A^\ta_{A}\cdot A^{\td}_D\bigg]
\label{secondorderone}
\end{align} 
where $k_P = k_A + k_B + k_C + k_D$ since $P=A\cup B\cup C\cup D$.
\para

Next, we collect all the terms in eqs.~(\ref{S1})-(\ref{S10}) 
with two powers of $g$ and in which the index $\mu$ labels a field:
\begin{align}
S^\mu_{(2,2)}
&=g^2
f^{\ta\td\tc}\Tc^{m n}
\sum_{P=A\cup B\cup C\cup D}
\left[ 
\overline{\tPsi}_{B}^m \gamma^\nu  \Psi^n_C  
+ (k^\nu_C-k^\nu_B)    
\tPhi^{m*}_B\Phi^n_C
\right]
A^\td_{\nu D} A^{\mu\ta}_A 			
\nn \\
&~~~~+g\Tc^{mn}\sum_{P=A\cup B\cup C}
\Big[
-\overline{\tPsi}_{B}^m \gamma_\nu  \Psi^n_C  
+ (k_{\nu B}-k_{\nu C})\tPhi^{m*}_B\Phi^n_C  \Big] H^{\nu\mu\tc}_A   
\nn\\
&~~~~+g\Tamn \sum_{P=A\cup B\cup C}
\left[ -2k^\nu_{A} 
\left(\tXi^{m*}_{\nu B}\Phi^n_C+\tPhi^{m*}_B\Xi^{n}_{\nu C}\right)
\right] A^{\mu\ta}_A 		
\nn\\
&~~~~+g\Tamn \sum_{P=A\cup B\cup C}
 2(k^\nu_B+k^\nu_C)
\left(\tXi^{\mu m*}_B\Phi^n_C+\tPhi^{m*}_B\Xi^{\mu n}_C\right)  
A^\ta_{\nu A} 			
\nn\\
&~~~~ + g^2 \sum_{P=A\cup B\cup C \cup D}
2 [T^\td,T^\ta]^{mn} 
\overline{\tPsi}_{B}^m \Aslash_D^\td \Psi^n_C A_A^{\mu\ta}   
\nn \\
&~~~~+g\Tamn \sum_{P=A\cup B\cup C}
\left( -2k^\nu_C\tXi^{\mu m*}_B\Phi^n_C
-2k^\nu_B\tPhi^{m*}_B\Xi^{\mu n}_C\right)
A^\ta_{\nu A} 				
\nn\\
&~~~~+g^2\sum_{P=A\cup B\cup C\cup D}
\Big[ -2(T^\td T^\ta)^{mn}k^\nu_B-2(T^\ta T^\td)^{mn}k^\nu_C\Big]  
\tPhi^{m*}_B\Phi^n_CA^\td_{\nu D}A^{\mu\ta}_A \,.    
\end{align}

We use \eqn{defH} and simplify the expression to obtain
\begin{align}
S^\mu_{(2,2)}
&=g^2
f^{\ta\td\tc}\Tc^{m n}
\sum_{P=A\cup B\cup C\cup D}
\Big[
2\overline{\tPsi}_{B}^m \gamma^\nu  \Psi^n_C 
+ 2(k^\nu_C-k^\nu_B) 
\tPhi^{m*}_B\Phi^n_C
\Big] 
A^\td_{\nu D}A^{\mu\ta}_A 
\nn\\[2mm]
&~~~~+g\Tamn \sum_{P=A\cup B\cup C}
\left[-2k^\nu_{A}
\left(\tXi^{m*}_{\nu B}\Phi^n_C+\tPhi^{m*}_B\Xi^{n}_{\nu C}\right)
\right]A^{\mu\ta}_A 
\nn \\
&~~~~+g\Td^{mn}\sum_{P=D\cup B\cup C}
\left[ 2k^\nu_B\tXi^{\mu m*}_B\Phi^n_C
+2k^\nu_C\tPhi^{m*}_B\Xi^{\mu n}_C\right]A^\td_{\nu D} \,.
\nn \\[2mm]
&~~~~ + g^2 \sum_{P=A\cup B\cup C \cup D}
2 [T^\td,T^\ta]^{mn} 
\overline{\tPsi}_{B}^m \Aslash_D^\td \Psi^n_C A_A^{\mu\ta}  
\nn \\[2mm]
&~~~~+ g^2\sum_{P=A\cup B\cup C\cup D}
\Big[-2(T^\td T^\ta)^{mn}k^\nu_B-2(T^\ta T^\td)^{mn}k^\nu_C\Big]
\tPhi^{m*}_B\Phi^n_CA^\td_{\nu D}A^{\mu\ta}_A  \,.
\end{align}

Then we apply \eqns{defXi}{defXitilde} to obtain
\begin{align}
S^\mu_{(2,2)}
&=
g^2 f^{\ta\td\tc}\Tc^{m n}
\sum_{P=A\cup B\cup C\cup D}
\Big[
2\overline{\tPsi}_{B}^m \gamma^\nu  \Psi^n_C 
+ 2(k^\nu_C-k^\nu_B) 
\tPhi^{m*}_B\Phi^n_C
\Big] 
A^\td_{\nu D}A^{\mu\ta}_A 
\nn\\
&~~~ +g^2\sum_{P=A\cup B\cup C\cup D}
\Big[
-2k^\nu_A (T^\td T^\ta)^{mn} - 2k^\nu_A (T^\ta T^\td)^{mn}
\nn \\
&~~~~ \hspace{30mm} 
+2(k^\nu_A+k^\nu_B)  (T^\ta T^\td)^{mn} 
+2(k^\nu_A+k^\nu_C)(T^\td T^\ta)^{mn}
\nn \\[2mm]
&~~~~ \hspace{30mm} 
-2k^\nu_B(T^\td T^\ta)^{mn} 
-2k^\nu_C(T^\ta T^\td)^{mn}
\Big]  \tPhi^{m*}_B\Phi^n_CA^\td_{\nu D}A^{\mu\ta}_A \nn
\nn\\
&~~~~ + g^2 \sum_{P=A\cup B\cup C \cup D}
2 [T^\td,T^\ta]^{mn} 
\overline{\tPsi}_{B}^m \Aslash_D^\td \Psi^n_C A_A^{\mu\ta}  
\nn \\[2mm]
&=
g^2 \sum_{P=A\cup B\cup C\cup D}
\Big[
2 f^{\ta\td\tc}\Tc^{m n}
+ 2 [T^\td,T^\ta]^{mn}  \Big]
\overline{\tPsi}_{B}^m \Aslash_D^\td \Psi^n_C  A_A^{\mu\ta}  
\\
&~~~+
g^2\sum_{P=A\cup B\cup C\cup D}
\Big[ 2(k^\nu_C-k^\nu_B)f^{\ta\td\tc}\Tc^{m n}
 + 2(k^\nu_C-k^\nu_B)[T^\td,T^\ta]^{mn}\Big]
\tPhi^{m*}_B\Phi^n_CA^\td_{\nu D}A^{\mu\ta}_A \,.
\nn 
\end{align}

Applying the commutation relations (\ref{commutationrelations}),
we see that this expression vanishes identically
\begin{align}
S^\mu_{(2,2)} 
&=0 \,.
\label{secondordertwo}
\end{align}

Finally we collect all the terms in eqs.~(\ref{S1})-(\ref{S10}) 
that have three powers of $g$,
relabeling some of the dummy indices
\begin{align}
S_{(3)}^\mu
=& g^2\sum_{P=A\cup B\cup C\cup D}
f^{\ta\td\tg}(T^\tg)^{m n}
\left(\tXi^{\nu m*}_B\Phi^n_C +\tPhi^{m*}_B\Xi^{\nu m}_C\right)  
A^\td_{\nu D}A^{\mu\ta}_A		
\nn\\
&
+g \sum_{P=G\cup B\cup C}
(T^\tg)^{mn}\left(-\tXi^{m*}_{\nu B}\Phi^n_C
-\tPhi^{m*}_B\Xi^{n}_{\nu C}\right)H^{\nu\mu\tg}_G  
\nn \\
&
+g\sum_{P=D\cup B\cup C}
\Td^{mn}\left(-\tXi^{\mu m*}_B\Xi^{\nu n}_C
+\tXi^{\nu m*}_B\Xi^{\mu n}_C\right)A^\td_{\nu D}  
\nn\\
&+g^2\sum_{P=A\cup B\cup C\cup D}
\left[
(T^\td T^\ta)^{mn}\tXi^{\nu m*}_B\Phi^n_C
-(T^\ta T^\td)^{mn}\tPhi^{m*}_B\Xi^{\nu n}_C\right]
A^\td_{\nu D}A^{\mu\ta}_A \,.    
\end{align}

Using the definitions (\ref{defH})-(\ref{defXitilde}), this becomes
\begin{align}
S_{(3)}^\mu
=g^3\sum_{P=A\cup B\cup C\cup D\cup E}
\Big\{
&
f^{\ta\td\tg}\left[ (T^\te T^\tg)^{mn}+(T^\tg T^\te)^{mn}\right] 
\nn \\
+&f^{\tg\te\ta}\left[ -(T^\td T^\tg)^{mn}-(T^\tg T^\td)^{mn}\right] 
\nn \\[3mm]
-&(T^\ta T^\td T^\te)^{mn} 
+(T^\te T^\td T^\ta)^{mn}
\nn \\[2mm]
+&(T^\te T^\td T^\ta)^{mn}
-(T^\ta T^\td T^\te)^{mn}
\Big\}
\tPhi^{m*}_B\Phi^n_C A^\td_{\nu D} A^{\nu\te}_E A^{\mu\ta}_A \,.
\end{align}

We now use the commutation relations (\ref{commutationrelations})
to eliminate the structure constants,
obtaining 
\begin{align}
S_{(3)}^\mu
&=g^3\sum_{P=A\cup B\cup C\cup D\cup E}
(T^\te[T^\ta,T^\td]
+[T^\ta,T^\td]T^\te
-T^\td[T^\te,T^\ta]
-[T^\te,T^\ta]T^\td
\nn\\
& \hspace{45mm}  -2T^\ta T^\td T^\te
+2T^\te T^\td T^\ta)^{mn}
\tPhi^{m*}_B\Phi^n_CA^\td_{D} \cdot A^{\te}_EA^{\mu\ta}_A 
\nn\\[3mm]
&=g^3\sum_{P=A\cup B\cup C\cup D\cup E}
(
T^\te T^\td T^\ta
-T^\ta T^\td T^\te
-T^\td T^\te T^\ta
+T^\ta T^\te T^\td
)^{mn} 
\tPhi^{m*}_B \Phi^n_C A^\td_{D} \cdot A^{\te}_E A^{\mu\ta}_A  \,.
\end{align}

Because the term in brackets is antisymmetric 
under $\td\leftrightarrow\te$ whereas $A^{\td}_D \cdot A_{E}^\te$ 
is symmetric (after summing over words),
this expression vanishes identically
\begin{align}
S^\mu_{(3)} &=0 \,.
\label{thirdorder}
\end{align}

Finally we now collect all of the results
(\ref{SYM}), (\ref{firstorder}), (\ref{secondorderone}), 
(\ref{secondordertwo}), and (\ref{thirdorder}) to obtain
\begin{align}
S^\mu 
&= \SYM + S_{(1)}^\mu + S_{(2,1)}^\mu + S_{(2,2)}^\mu + S_{(3)}^\mu
\nn\\
&=
k_P^\mu 
\bigg[ 
g f^{\ta\tb\tc} \sum_{P=A\cup B\cup C}
 k_A \cdot A_C^\tc \ A_A^\ta \cdot A_B^\tb
-
\half g^2  
f^{\ta\tb\te}
f^{\te\tc\td}
\sum_{P=A\cup B\cup C \cup D}
A_A^\ta \cdot A_C^\tc
\ A_B^\tb \cdot A_D^\td 
\nn\\
&
~~~~~~~
+ g \Tamn \sum_{P=A\cup B\cup C}
(k_B-k_C) \cdot A^\ta_{A} \tPhi^{m*}_B\Phi^n_C
- g \Tamn \sum_{P=A\cup B\cup C}
\overline{\tPsi}_{B}^m \Aslash_A^\ta \Psi^n_C 
\nn\\
&
~~~~~~~
- 2g^2 
(T^\ta T^\td)^{mn}
\sum_{P=A\cup B\cup C\cup D}
\tPhi^{m*}_B\Phi^n_C 
A^\ta_{A} \cdot A^{\td}_D \bigg] \,.
\end{align}

Recalling that $P=123...(n-1)$ and using momentum conservation
$\sum_{i=1}^n  k_i=0$, we have that $k_P=-k_n$. 
Then contracting $S^\mu_n$ with $\eps_{\mu n}$,
and using $\eps_n \cdot k_n =0$,
we finally obtain 
\begin{align}
\delta_n \cA_n=0 \,.
\end{align}

Naturally the demonstration is equivalent for any of the other gluons
in the amplitude, and therefore we have demonstrated the
invariance of the $n$-point amplitude under the color-factor shift 
associated with any of the gluons in the amplitude.
The BCJ relations immediately follow, as shown in sec.~\ref{sec:cfs}.

\section{Conclusions}
\setcounter{equation}{0}
\label{sec:concl}

In this paper, we have applied perturbiner methods to obtain a recursive
expression for tree-level amplitudes in Yang-Mills gauge theories
coupled to matter with an arbitrary number $n$ of external particles,
whether gluons or scalar or spinor matter particles.  We then employed
this expression to provide a streamlined proof that, for any $n$,
the amplitude is invariant under a color-factor shift associated with
each of the gluons in the amplitude, thus confirming the results of
ref.~\cite{Brown:2016mrh,Brown:2016hck}.  The strength of this recursive
approach is that it allows one to establish all orders in $n$ results
without having to explicitly evaluate any particular amplitude.
\para

Color-factor symmetry is a property of tree-level amplitudes closely
related to, but distinct from, color-kinematic duality.  The invariance of
the amplitude under a color-factor shift is a gauge-invariant statement
(whereas color-kinematic duality is only satisfied in a particular
generalized gauge), and can be used to directly infer the BCJ relations.
Discovering if and how color-factor symmetry extends to loop level 
remains an important goal.

\section*{Acknowledgments}
This material is based upon work supported by the
National Science Foundation
under Grant No.~PHY21-11943.

\vfil\break
\appendix

\section{Perturbiner computation of matter-antimatter annihilation}
\setcounter{equation}{0}
\label{sec:app}

In this appendix, we illustrate the use of the general expression 
developed at the end of sec.~\ref{sec:pert}, namely
\begin{align}
\cA_n
=
g \eps_{\mu n} \Bigg( f^{\ta_n\tb\tc}
\sum_{P=Q\cup{R}}A^\tb_{\nu{Q}}G^{\nu\mu\tc}_R
&+  (T^{\ta_n})^{mn} \sum_{P=Q \cup R}
\overline{\tPsi}_{Q}^m \gamma^\mu  \Psi^n_R
\nn\\
&
-  (T^{\ta_n})^{mn}\sum_{P=Q\cup{R}}
\left[\tTh^{\mu{m*}}_Q\Phi^n_R +\tPhi^{m*}_Q\Th^{\mu{n}}_R\right]
\Bigg)  
\label{npointampapp}
\end{align}

to compute the amplitude for matter-antimatter annihilation 
(for scalar or spinor matter fields) into two gluons.
(This was also done in ref.~\cite{Gomez:2020vat}.)
For the four-point amplitude we have $P=123$,
which can be divided into the following pairs
$Q \cup R$ of non-empty ordered words:
$ 12 \cup 3 $, $ 13 \cup 2 $, $23 \cup 1 $, 
$ 1 \cup 23 $, $ 2 \cup 13 $, and $3 \cup 12 $. 
\para

First we consider the scalar annihilation amplitude
$\cA_4 (\bar{\Phi}_1, \Phi_2, g_3, g_4)$.
The only terms in \eqn{npointampapp} that contribute to the amplitude 
are 
\begin{align}
\cA_4
=
g \eps_{\mu 4} \left(  f^{\ta_4\tb\tc}
\left[ 
A^\tb_{\nu{12}}G^{\nu\mu\tc}_3
+A^\tb_{\nu{3}}G^{\nu\mu\tc}_{12}
\right]
- (T^{\ta_4})^{mn}
\left[
 \tTh^{\mu{m*}}_{13}\Phi^n_2 
+\tPhi^{m*}_{13}\Th^{\mu{n}}_2
+\tTh^{\mu{m*}}_1\Phi^n_{23} 
+\tPhi^{m*}_1\Th^{\mu{n}}_{23}
\right]
\right)  \,.
\label{fourpointscalar}
\end{align}

The terms with single letter words are evaluated using 
\eqns{linear}{linearaux}
\begin{align}
A_3^{\mu \,\ta}  
&= \eps_3^\mu \delta^{\ta\ta_3} \,,
& 
G_3^{\nu\mu\,\ta} &= 
(-2 k_3^\nu \eps_3^\mu + k_3^\mu \eps_3^\nu )
\delta^{\ta\ta_3} \,,
\nn\\[2mm]
\tPhi_1^{n*} &= \tphi_1^* \delta^{nn_1}  \,,
&
\Phi_2^n &= \phi_2 \delta^{nn_2}  \,,
\nn \\[2mm]
\tTh_1^{\mu n *} &=   k_1^\mu \tphi_1^* \delta^{nn_1}  \,,
&
\Th_2^{\mu n} &= - k_2^\mu \phi_2 \delta^{nn_2}  
\label{linearfourpointscalar}
\end{align}

and all terms with double letter words in \eqn{fourpointscalar}
may be evaluated using eqs.~(\ref{kA})-(\ref{kThtilde})
\begin{align}
k_{12}^2 A_{12}^{\mu  \ta} 
&=   g (T^\ta)^{mn} 
\left[ -{\tTh}^{\mu m * }_{1} \Phi^n_2
- {\tPhi}^{m *}_{1}  \Th^{\mu n}_2 \right] \,,
\nn\\[2mm]
k_{12}^2 G_{12}^{\nu\mu  \ta}
&= 
 g (T^\ta)^{mn} 
\left[ 2 k^\nu_{12} 
(  {\tTh}^{\mu m*}_1 \Phi_2^n + {\tPhi}^{m*}_1 \Th_2^{\mu n} )
- k^\mu_{12} 
(  {\tTh}^{\nu m*}_1 \Phi_2^n + {\tPhi}^{m*}_1 \Th_2^{\nu n} ) \right] \,,
\nn\\[2mm]
(k_{23}^2 - \ms^2) \Phi_{23}^m 
&= g (T^\ta)^{mn} A_{\mu 3}^{\ta} \Ups^{\mu n}_2 \,,
\nn\\[2mm]
(k_{13}^2 - \ms^2) \tPhi_{13}^{m  *}
&= g (T^\ta)^{nm} \tUps^{\mu n *}_1  A_{\mu 3}^{\ta}  \,,
\nn\\[2mm]
(k_{23}^2 - \ms^2) \Th_{23}^{\mu m}
&= g (T^\ta)^{mn} 
\left[ -k^\mu_{23} A^{\ta}_{\nu 3} \Ups^{\nu n}_2
 - (k_{23}^2-\ms^2)  A^{\mu \ta}_{3} \Phi^n_2 \right] \,,
\nn\\[2mm]
(k_{13}^2 - \ms^2) \tTh_{13}^{\mu m *}
&= g (T^\ta)^{nm} 
\left[  k^\mu_{13} \tUps^{\nu n *}_1 A^{\ta}_{\nu 3} 
- (k_{13}^2-\ms^2) \tPhi^{n*}_1  A^{\mu \ta}_{3} \right]
\end{align}

which may in turn be evaluated using \eqn{linearfourpointscalar} together with
\begin{align}
\tUps_1^{\mu n *} &= 2k_1^\mu \tphi_1^* \delta^{nn_1}  \,,
&
\Ups_2^{\mu n} &= -2k_2^\mu \phi_2 \delta^{nn_2}  \,.
\end{align}

We also use $k_1^2=k_2^2=\ms^2$ and $k_3^2 = k_4^2 =0$ and
$\sum_{i=1}^4 k_i^\mu=0$ 
to evaluate the invariants
\begin{align}
k_{12}^2 = k_{34}^2 &= 2 k_3 \cdot k_4 \,, \nn\\[2mm]
k_{13}^2 - \ms^2 &= k_{24}^2 - \ms^2 = 2 k_2 \cdot k_4 \,, \nn\\[2mm]
k_{23}^2 - \ms^2 &= k_{14}^2 -\ms^2 = 2 k_1 \cdot k_4 \,.
\label{invariants}
\end{align}

Putting everything together, we obtain
\begin{align}
\cA_4
& = 
g^2 \tphi_{1}^{ *} \phi_2 
\left[  
{f^{\ta_4 \ta_3 \te} (T^\te)^{n_1 n_2} \over 2 k_3 \cdot k_4}
n_{(3)}
+ 
{ (T^{\ta_3} T^{\ta_4})^{n_1 n_2} \over 2 k_2 \cdot k_4} 
n_{(2)} 
- 
{(T^{\ta_4} T^{\ta_3})^{n_1 n_2} \over 2 k_1 \cdot k_4} 
n_{(1)} 
\right] 
\label{fourpointperturbiner}
\end{align}

where the kinematic numerators $n_{(i)}$ are
given by
\begin{align}
n_{(3)} 
& = 
 \eps_4 \cdot k_1 \,( \eps_3 \cdot k_1 + 3 \eps_3 \cdot k_2)
+\eps_4 \cdot k_2 \,( -3 \eps_3 \cdot k_1 - \eps_3 \cdot k_2)
\nn\\
&
~~~+\eps_4 \cdot k_3 \,( \eps_3 \cdot k_1 - \eps_3 \cdot k_2)
+\eps_4 \cdot \eps_3 \,(- 2k_1 \cdot k_3 + 2 k_2 \cdot k_3) \nn
\nn \\
& = 
 \eps_4 \cdot k_1 \,(  4 \eps_3 \cdot k_2)
+\eps_4 \cdot k_2 \,( -4 \eps_3 \cdot k_1)
+\eps_4 \cdot \eps_3 \,(-2 k_2 \cdot k_4 + 2 k_1 \cdot k_4) \,,
\nn\\[3mm]
n_{(2)} 
& = 
 \eps_4 \cdot k_1 \,( -2 \eps_3 \cdot k_1 )
+\eps_4 \cdot k_2 \,( 2 \eps_3 \cdot k_1 )
+\eps_4 \cdot k_3 \,( -2 \eps_3 \cdot k_1 )
+\eps_4 \cdot \eps_3 \,(2 k_1 \cdot k_3)
\nn \\
& = 
 \eps_4 \cdot k_2 \,( 4 \eps_3 \cdot k_1 )
+\eps_4 \cdot \eps_3 \,(2 k_2 \cdot k_4) \,,
\nn\\[3mm]
n_{(1)} 
& =  
 \eps_4 \cdot k_1 \,( -2 \eps_3 \cdot k_2 )
+\eps_4 \cdot k_2 \,( 2 \eps_3 \cdot k_2 )
+\eps_4 \cdot k_3 \,( 2 \eps_3 \cdot k_2 )
+\eps_4 \cdot \eps_3 \,(-2 k_1 \cdot k_4)
\nn\\
& = 
 \eps_4 \cdot k_1 \,( -4 \eps_3 \cdot k_2 )
+\eps_4 \cdot \eps_3 \,(-2 k_1 \cdot k_4)
\end{align}

where we have used 
$\sum_{k=1}^4 k^\mu_i = 0$ and $\eps_3 \cdot k_3 = 0$
to simplify the expressions.
Setting $\tilde\phi_1 =\phi_2=1$, this result  precisely
matches \eqn{fourpointamp} obtained using Feynman diagrams.
The kinematic Jacobi identity 
$n_{(1)} +n_{(2)} + n_{(3)}  = 0$
is manifestly satisfied \cite{Zhu:1980sz,Goebel:1980es}.
\para

Next we consider the spinor annihilation amplitude
$\cA_4 (\bar{\Psi}_1, \Psi_2, g_3, g_4)$.
The only terms in \eqn{npointampapp} that contribute to the amplitude 
are 
\begin{align}
\cA_4
=
g \eps_{\mu 4} \left(  f^{\ta_4\tb\tc}
\left[ 
A^\tb_{\nu{12}}G^{\nu\mu\tc}_3
+A^\tb_{\nu{3}}G^{\nu\mu\tc}_{12}
\right]
+ (T^{\ta_4})^{mn}
\left[
\overline{\tPsi}_{13}^m \gamma^\mu  \Psi^n_2
+
\overline{\tPsi}_{1}^m \gamma^\mu  \Psi^n_{23}
\right]
\right)  \,.
\label{fourpointspinor}
\end{align}

The terms with single letter words are evaluated using \eqn{linear}
\begin{align}
A_3^{\mu \,\ta}  
&= \eps_3^\mu \delta^{\ta\ta_3} \,,
& 
G_3^{\nu\mu\,\ta} &= 
(-2 k_3^\nu \eps_3^\mu + k_3^\mu \eps_3^\nu )
\delta^{\ta\ta_3} \,,
\nn\\[2mm]
\overline{\tPsi}_1^n &= \overline{\tpsi}_1 \delta^{nn_1}  \,,
&
\Psi_2^{n} &= \psi_2  \delta^{nn_2}  
\label{linearfourpointspinor}
\end{align}

and all terms with double letter words in \eqn{fourpointspinor}
may be evaluated using eqs.~(\ref{kA})-(\ref{kPsi})
\begin{align}
k_{12}^2 A_{12}^{\mu  \ta} 
&=    g (T^\ta)^{mn} 
\overline{\tPsi}_{1}^m \gamma^\mu  \Psi^n_2
\,, \nn\\[2mm]
k_{12}^2 G_{12}^{\nu\mu  \ta}
&= 
g (T^\ta)^{mn} 
\left[
-2 k^\nu_{12} 
(  \overline{\tPsi}^m_1 \gamma^\mu \Psi_2^n )
+ k^\mu_{12} 
(  \overline{\tPsi}^m_1 \gamma^\nu \Psi_2^n )
\right]
\,,\nn\\[2mm]
( \kslash_{23} - \mf ) \Psi_{23}^m 
&= - g (T^\ta)^{mn} \Aslash_3^{ \ta} \Psi_2^n 
\,,\nn\\[2mm]
\overline{\tPsi}_{13}^m 
( \kslash_{13} + \mf ) 
&=   g (T^\ta)^{nm} \overline{\tPsi}_1^n \Aslash_3^{ \ta}  \,.
\end{align}

Putting everything together, using \eqn{invariants}, we obtain
\begin{align}
\cA_4
& = 
g^2 
\left[  
{f^{\ta_4 \ta_3 \te} (T^\te)^{n_1 n_2} \over 2 k_3 \cdot k_4}
n_{(3)}
+ 
{ (T^{\ta_3} T^{\ta_4})^{n_1 n_2} \over 2 k_2 \cdot k_4} 
n_{(2)} 
- 
{(T^{\ta_4} T^{\ta_3})^{n_1 n_2} \over 2 k_1 \cdot k_4} 
n_{(1)} 
\right] 
\end{align}

where the kinematic numerators $n_{(i)}$ are given by
\begin{align}
n_{(3)} 
& = 
   2 \eps_4 \cdot \eps_3 ( \overline{\tpsi}_{1} \kslash_3 \psi_2 )
- \eps_4 \cdot k_3  ( \overline{\tpsi}_{1} \epsslash_3  \psi_2 )
- 2 \eps_3 \cdot k_{12} (  \overline{\tpsi}_1 \epsslash_4  \psi_2 )
+ \eps_4  \cdot k_{12} (  \overline{\tpsi}_1 \epsslash_3  \psi_2 )
\nn\\
& = 
  2 \eps_4 \cdot \eps_3 ( \overline{\tpsi}_{1} \kslash_3 \psi_2 )
- 2 \eps_4  \cdot k_3 (  \overline{\tpsi}_1 \epsslash_3  \psi_2 )
+ 2 \eps_3 \cdot k_4 (  \overline{\tpsi}_1 \epsslash_4  \psi_2 ) \,,
\nn\\[2mm]
n_{(2)} 
& =   \overline{\tpsi}_1 \epsslash_3  (\kslash_{13} - m ) \epsslash_4 \psi_2
\nn\\
& =  - \overline{\tpsi}_1 (\kslash_{13} + m )\epsslash_3   \epsslash_4 \psi_2
    +  2 \eps_3 \cdot k_1 (\overline{\tpsi}_1  \epsslash_4 \psi_2)
\nn\\
& = -  \overline{\tpsi}_1 (\kslash_{1} + m )\epsslash_3   \epsslash_4 \psi_2
    +  \overline{\tpsi}_1 \kslash_{3} \epsslash_4   \epsslash_3 \psi_2
    -  2 \eps_4 \cdot  \eps_3 (\overline{\tpsi}_1 \kslash_{3}  \psi_2)
    +  2 \eps_3 \cdot k_1 (\overline{\tpsi}_1  \epsslash_4 \psi_2)
\nn\\
& =    \overline{\tpsi}_1 \kslash_{3} \epsslash_4   \epsslash_3 \psi_2
    -  2 \eps_4 \cdot  \eps_3 (\overline{\tpsi}_1 \kslash_{3}  \psi_2)
    +  2 \eps_3 \cdot k_1 (\overline{\tpsi}_1  \epsslash_4 \psi_2) \,,
\nn\\[2mm]
n_{(1)} 
& =    \overline{\tpsi}_1 \epsslash_4 ( \kslash_{23} + m )  \epsslash_3 \psi_2  
\nn\\
& =    2 \eps_4 \cdot k_3 (\overline{\tpsi}_1  \epsslash_3 \psi_2)
    -   \overline{\tpsi}_1 \kslash_{3} \epsslash_4 \epsslash_3 \psi_2  
    +   2 \eps_3 \cdot k_2 (\overline{\tpsi}_1 \epsslash_4  \psi_2)
    -   \overline{\tpsi}_1 \epsslash_4 \epsslash_3 ( \kslash_{2} - m )  \psi_2  
\nn\\
& =    2 \eps_4 \cdot k_3 (\overline{\tpsi}_1  \epsslash_3 \psi_2)
    -   \overline{\tpsi}_1 \kslash_{3} \epsslash_4 \epsslash_3 \psi_2  
    +   2 \eps_3 \cdot k_2 (\overline{\tpsi}_1 \epsslash_4  \psi_2)
\end{align}

where we have used 
$\sum_{k=1}^4 k^\mu_i = 0$, 
$\eps_3 \cdot k_3 = 0$ ,
and 
$\{\gamma^\mu, \gamma^\nu \} = 2 \eta^{\mu\nu} $,
as well as \eqn{linear} to simplify the expressions.
These expressions agree with \eqn{spinornumerators}
upon setting $\psi_2 = u_2$ and $\tpsi_1 = v_1$.
One also easily verifies the kinematic 
Jacobi identity \cite{Zhu:1980sz,Goebel:1980es}
\begin{align}
n_{(1)} +n_{(2)} + n_{(3)} 
& = 
  2 \eps_3 \cdot (k_1 + k_2 + k_4) \overline{\tpsi}_1  \epsslash_4 \psi_2
 = 0 \,.
\end{align}

\vfil\break


\end{document}